\documentclass[pra,aps,twocolumn,10pt,floatfix,nofootinbib,superscriptaddress]{revtex4-1}

\pdfoutput=1

\usepackage[english]{babel}
\usepackage{graphicx}
\usepackage{dcolumn}
\usepackage{bm}
\usepackage{physics}
\usepackage{placeins}
\usepackage{xcolor}
\usepackage{amsmath}
\usepackage{amssymb}
\usepackage{bbold}
\usepackage{txfonts}
\usepackage{mathrsfs}
\usepackage{multirow}
\usepackage[version=4]{mhchem}   
\usepackage[colorlinks=true,citecolor=blue,linkcolor=red,linktocpage=true]{hyperref}

        \definecolor{AAcolor}{rgb}{0.7,0.1,0.4}

		\newcommand{\e}[1]{\begin{align}{#1}\end{align}}	
		
		\newcommand{\lin}{\notag \\}
		
		\newcommand{\f}[2]{\frac{#1}{#2}}

		\newcommand{\la}[1]{\label{#1}}

		\newcommand{\q}[1]{Eq.~(\ref{#1})}
		\newcommand{\s}[1]{Sec.\ \ref{#1}}
		\newcommand{\fig}[1]{Fig.~\ref{#1}}

		\newcommand{\ocite}[1]{Ref.\ [\onlinecite{#1}]}

	\newcommand\as{\;\;\;\;}			

		\newcommand{\sgn}{\text{sgn}}

		\newcommand{\eq}{=&\;}

		\newcommand{\ri}{\rightarrow}

        \newcommand{\calh}{\mathscr{H}} 

        \newcommand{\bkp}{\bk_{\perp}}
        
        \newcommand{\br}{\boldsymbol{r}}
        \newcommand{\bR}{\boldsymbol{R}}
        \newcommand{\bze}{\boldsymbol{0}}
        \newcommand{\bk}{{\boldsymbol{k}}}
        
        \newcommand{\bA}{\boldsymbol{A}}
        \newcommand{\bG}{\boldsymbol{G}}
        
        \newcommand{\nabk}{\nabla_\bk}
        \newcommand{\tk}{\textrm{K}}
        \newcommand{\tkpr}{{\textrm{K}'}}
        \newcommand{\tm}{\textrm{M}}

        \newcommand{\ts}{\textrm{S}}
        \newcommand{\bw}{\boldsymbol{w}}

        \newcommand{\bsigma}{\bm{\sigma}}
        \def\bs#1{\boldsymbol{#1}}	 

        \newcommand{\noi}[1]{\noindent (#1)}

	\newcounter{subeqn} %
	\makeatletter
	\@addtoreset{subeqn}{equation}
	\makeatother

	\newcommand*\mcup{\mathbin{\mathpalette\mcupinn\relax}}
\newcommand*\mcupinn[2]{\vcenter{\hbox{$\mathsurround=0pt
  \ifx\displaystyle#1\textstyle\else#1\fi\bigcup$}}}

   \def\EBR#1#2#3#{${#1}\uparrow\mathsf{G}({#2})\;@{#3}$}
   \def\allpoints{$\Gamma$--K, $\Gamma$--M, K--M}
   \def\allprimed{$\Gamma$--K, $\Gamma$--M, K--M, $\Gamma$--K$'$, K$'$--M, K--K$'$}
   \def\GKKprime{$\Gamma$--K, $\Gamma$--K$'$, K--K$'$}
   \def\adjoin{$\quad$and$\quad$}
   
\usepackage[normalem]{ulem} 
\usepackage{soul}                   
\definecolor{TB}{rgb}{0,0.72,0.92}

\definecolor{AN}{rgb}{0.1,0.7,0.4}

\definecolor{changes}{rgb}{0.1,0.6,0.1}

\newcounter{myfigure}
\makeatletter
\@addtoreset{figure}{myfigure}
\makeatother
    
\begin{document} 

   \title{Multicellularity of delicate topological insulators}
    \author{Aleksandra Nelson}
    \email{anelson@physik.uzh.ch}
    \affiliation{Department of Physics, University of Zurich, Winterthurerstrasse 190, 8057 Zurich, Switzerland}
    \author{Titus Neupert}
    \affiliation{Department of Physics, University of Zurich, Winterthurerstrasse 190, 8057 Zurich, Switzerland}
    \author{Tom\'{a}\v{s} Bzdu\v{s}ek}
    \affiliation{Condensed Matter Theory Group, Paul Scherrer Institute, 5232 Villigen PSI, Switzerland}
    \affiliation{Department of Physics, University of Zurich, Winterthurerstrasse 190, 8057 Zurich, Switzerland}
    \author{A. Alexandradinata}
    \email{aalexan7@illinois.edu}
    \affiliation{Department of Physics, University of Illinois at Urbana-Champaign, Urbana, Illinois 61801-2918, USA}%
   \begin{abstract}
   Being Wannierizable is not the end of the story for topological insulators. We introduce a family of topological insulators that would be considered trivial in the paradigm set by the tenfold way, topological quantum chemistry, and the method of symmetry-based indicators. Despite having a symmetric, exponentially-localized Wannier representation, each Wannier function cannot be completely localized to a single primitive unit cell in the bulk. Such \textit{multicellular topology} is shown to be neither stable, nor fragile, but \textit{delicate}, i.e., the topology can be nullified by adding trivial bands to either valence or conduction band. 
   \end{abstract}
   \date{\today}
   \maketitle

\emph{Introduction.---} Two themes have indelibly shaped the paradigm of topological insulators (TIs), and couched how topological properties are discussed, modelled, and measured. The first is the notion of stability of TIs, and the second involves the various obstructions to forming a real-space Wannier-function (WF) representation of the valence band~\cite{Thouless_wannierfunctions,thonhauser_cherninsulator,Brouder_explocWFs,alexey_wannierrepZ2TI,Maryam_wanniercentersheets,budich_wannierfunctions,TBO_JHAA,read_compactwannier}. This work describes an extension and fine-graining of both themes, and introduces a novel family of TIs that would be considered unstable and unobstructed according to the presently-held paradigm. 

\begin{figure}[b!]
    \centering
    \includegraphics[width=0.48\textwidth]{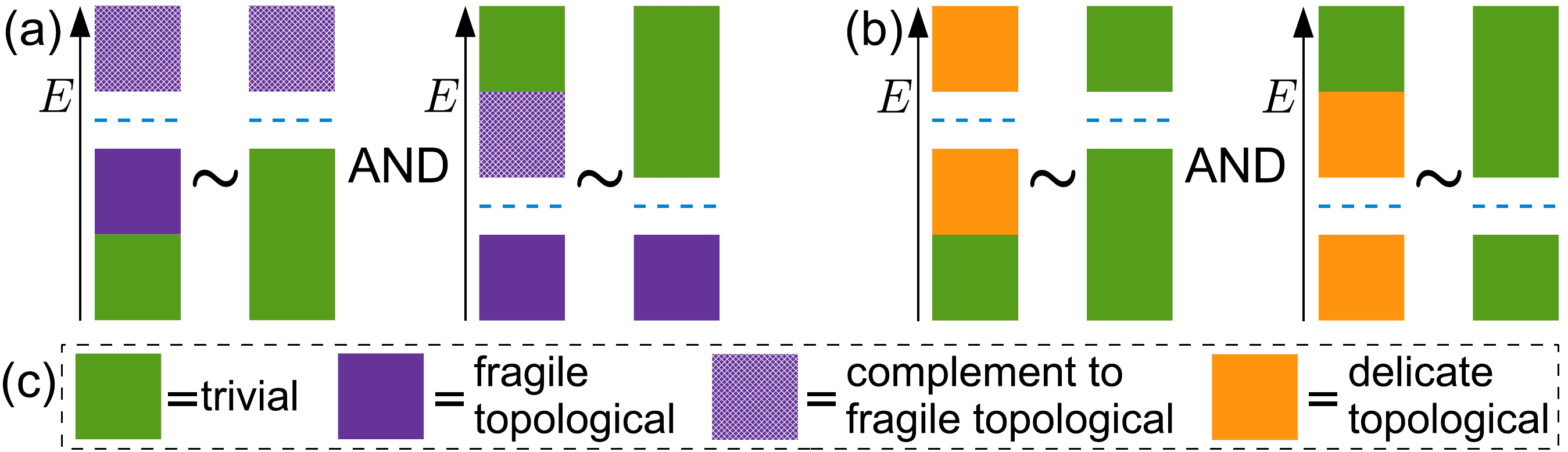}
        \caption{Topological stability of (a) fragile vs.~(b) delicate topology. The blocks [colored according to legend in (c)] below/above the Fermi energy (dashed blue lines) are valence/conduction bands, with addition indicated by stacking and topological equivalence by `$\sim$'. In panel (a) the complement to fragile topological bands could be fragile topological, obstructed atomic limit, or fully trivial.} \label{fig:fragile-vs-delicate}
\end{figure}

The strongest form of stability is the notion of stable equivalence introduced by $K$-theory~\cite{read_compactwannier,kitaev_periodictable,FreedMoore_twistedequivariantmatter,bandcombinatorics_kruthoff,shiozaki_review}, where the bulk/surface topological invariant of a valence subspace is immune to addition of trivial bands. The intermediate notion  of fragility means that the topological property can be nullified by adding trivial bands to the valence subspace, but not to the conduction subspace~\cite{po_fragile,bouhon_wilsonloopapproach,bradlyn_disconnectedEBR,else_fragile,zhida_fragileaffinemonoid,Bouhon:2020,crystalsplit_AAJHWCLL} [Fig.~\ref{fig:fragile-vs-delicate}(a)]. A~distinct notion that we introduce here is \textit{delicate topology}, where  the topological property can be nullified by adding trivial bands to either valence or conduction subspace~[Fig.~\ref{fig:fragile-vs-delicate}(b)]. For symmetry-protected delicate topology, nullification occurs only by adding trivial bands of certain symmetry representations. 

\begin{figure}[b!]
    \centering
    \includegraphics[width=0.48\textwidth]{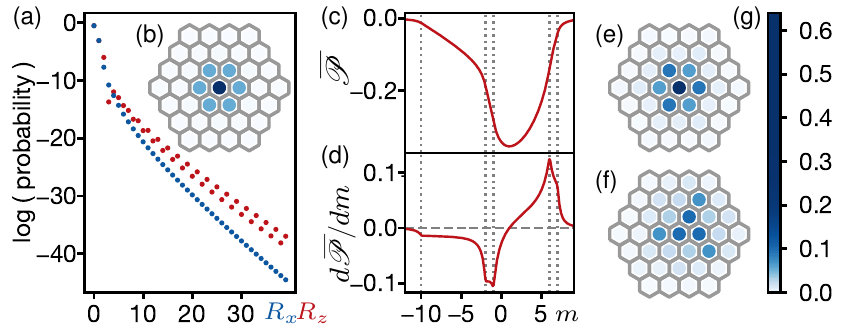}
    \caption{Characterization of bulk (panels a-d) and surface (e,f) Wannier functions for the delicate topological insulator modelled by Eq. (1) with $m{=}{-}6,\chi{=}{-}1$.   (a) illustrates the exponential decay parallel (orthogonal) to the rotation axis, as indicated by $R_z$ ($R_x$). (b,e,f) utilize the color bar in (g) to illustrate the rotation-symmetric probability distribution -- projected onto the rotation-invariant plane. (c--d) 
    The total polarization in $z$ direction
    is non-analytic at topological-phase transitions indicated by dashed lines.}  \label{fig:wannier}
\end{figure}

Many authors have proposed a useful definition of a trivial band to be its possession of an exponentially-localized WF representation respecting the crystallographic spacetime symmetries~\cite{bandcombinatorics_kruthoff,shiozaki_review,TQC,cano_buildingblocksTQC,Po_symmetryindicators,nogo_AAJH,crystalsplit_AAJHWCLL}. By this definition, all stably-equivalent and fragile TIs present an obstruction to a WF representation. It has been further argued through equivariant vector bundle theory that such Wannier obstructions represent a robust property of a valence subspace summed with an \textit{arbitrary} conduction subspace~\cite{crystalsplit_AAJHWCLL}, and therefore such obstruction cannot exist for delicate topological insulators. Here, we introduce a distinct class of obstructions that prevents WFs from being completely localized to a single, primitive unit cell -- we call this \textit{multicellular topology} [Fig.~\ref{fig:wannier}(a--c)]. Conversely, we adopt a distinct notion of triviality, namely that symmetry-respecting WFs exist and can be confined to a single cell by a continuous, adiabatic deformation of the Hamiltonian -- \emph{unicellularity}.

The notions of delicate and multicellular topology are distinct and a priori need not come together in any specific realization of a TI. This work aims to open the debate by presenting a concrete family of tight-binding models which simultaneously manifests both types of topology, and sets the stage for future realizations and discoveries.

\emph{Returning Thouless pump.---} We begin by introducing a class of tight-binding models in three spatial dimensions, which exhibit both symmetry-protected delicacy and multicellularity. The relevant symmetry is an $n$-fold rotation symmetry $C_n$ about the $z$ axis. The tight-binding Hilbert space is given by an orthonormal set of WFs $\{\varphi_{j,\bR}\}_{\bR\in \textrm{BL},j=1\ldots \mathcal{C}{+}\mathcal{V}}$ over   the Bravais lattice ($\textrm{BL}$), which satisfy the \textit{uniaxial condition}, i.e., that all independent WFs within a representative, primitive unit cell are centered on the same rotational axis, and individually form one-dimensional representations of $C_n$. (This simplifying assumption holds only for the subclass of multicellular TIs studied here.)  This allows to decompose the Hilbert space as $\calh[\varphi]{=}{\oplus_{\ell=0}^{n-1}}\calh_\ell[\varphi]$, where the summands are distinguished by the $n$ possible angular momenta $\ell$, with corresponding rotation eigenvalues $e^{i2\pi\ell/n}$. 

We further assume the valence (resp.~conduction) bands can be spanned by exponentially-localized WFs $\{W^v_{j,\bR}\}_{\bR\in \textrm{BL},j=1\ldots \mathcal{V}}$ (resp.~$\{W^c_{j,\bR}\}_{\bR\in\textrm{BL},j=1\ldots \mathcal{C}}$). Though generally distinct from $\{\varphi_{j,\bR}\}$,  we demand that $\{W^{c/v}_{j,\bR}\}$ also satisfy the uniaxial condition, and additionally satisfy the \textit{mutually-disjoint condition} -- that any representation appearing in the valence subspace ($\calh[W^v]{=}{\oplus_{\ell_v}}\calh_{\ell_v}[W^v]$) cannot appear in the conduction subspace ($\calh[W^c]{=}{\oplus_{\ell_c}}\calh_{\ell_c}[W^c]$, with $\ell_{v(c)}$ disjoint). 

The uniaxial condition on exponentially-localized WFs implies that both conduction and valence bands are band representations~\cite{Zak_bandrepresentations,Bacry_bandrepresentations}, making the system trivial from the viewpoints of topological quantum chemistry~\cite{TQC} and symmetry-based indicators~\cite{Po_symmetryindicators}. A band representation also precludes a nontrivial first Chern class~\cite{nogo_AAJH,crystalsplit_AAJHWCLL}, making the model trivial in the tenfold way~\cite{kitaev_periodictable,schnyder_classify3DTIandTSC,schnyder_classifyTIandTSC}.  Nevertheless, we find that the mutually-disjoint condition allows for a type of symmetry-protected multicellularity, where the WFs necessarily extend -- \emph{beyond one unit cell} -- in the direction of the rotation axis. 

The multicellularity manifests in the discrete spectrum of the projected position operator $P\hat{z}P$~\cite{marzari1997,AA_wilsonloopinversion}, with $P$ projecting to the bulk valence band. Since $P\hat{z}P$ is invariant under translations perpendicular to the rotation axis, each  eigenvalue of $P\hat{z}P$ forms a band over the two-dimensional (2D) \emph{reduced Brillouin zone}, $\textrm{rBZ}{\ni}\bkp{=}(k_x,k_y)$. Under translation along the rotation axis by a lattice period (set to one),  $P\hat{z}P{\ri} P(\hat{z}{+}1)P$, hence each eigenvalue belongs to an infinitely-extended Wannier-Stark ladder~\cite{wannier_starkladder}, and the full spectrum comprises $\mathcal{V}$ such ladders which are non-degenerate at generic $\bkp$~\cite{TBO_JHAA}. We pick one representative eigenvalue from each ladder, and define their sum (modulo integer) to be the (charge) polarization $\mathscr{P}(\bkp)$, in accordance with the geometric theory of polarization~\cite{zak_berryphase,kingsmith_polarization}.   

Since distinct rotational representations cannot mix at $C_n$-invariant points ($\bk'_{\perp}{\equiv} C_n\bk'_{\perp}$), the polarization can be decomposed into a sum of polarizations in each angular-momentum sector:  $\mathscr{P}(\bkp'){=}{\sum_{\ell_v}}\mathscr{P}_{\ell_v}(\bkp')$. This non-mixing, combined with the mutually-disjoint condition, implies an identity between symmetry-decomposed  Hilbert spaces $\calh_{\ell_v}[W^v]|_{\bk'_{\perp}}{=}\calh_{\ell_v}[\varphi]|_{\bk'_{\perp}}$ when restricted to any $C_n$-invariant wavevector. It follows that the polarization $\mathscr{P}_{\ell_v}(\bkp')$ equals, modulo integer, to the polarization of the basis WFs in the spin sector $\ell_v$; the latter quantity is $\bkp'$-independent because any tight-binding basis function has support only on a single lattice site. Therefore, modulo integer, $\mathscr{P}_{\ell_v}(\bkp')$ is independent of $\bkp'$, and hence also $\mathscr{P}(\bkp')$. If $\mathscr{P}(\bkp)$ is continuously defined over $\textrm{rBZ}$ with multiple $C_n$-invariant points, the difference $\Delta\mathscr{P}_{\bkp'\bkp''}{:}{=}\mathscr{P}(\bkp''){-}\mathscr{P}(\bkp')$ between any pair of these points is quantized to integers. $\Delta\mathscr{P}_{\bkp'\bkp''}{=}\mu$ implies a Thouless pump~\cite{thouless_pump} of $\mu$ electron charges over one half-period of the $\textrm{rBZ}$ (connecting $\bkp'$ and $\bkp''$); the triviality of the first Chern class ensures that this charge is reversed in the second half-period. Such a \emph{returning Thouless pump} (RTP) guarantees that: (i) the Hamiltonian cannot be adiabatically deformed to be $\bk$-independent (having no hopping elements in real space), and (ii) at least one WF must extend over \emph{multiple unit cells} in the direction of the rotation axis (see SM~\cite{supp}).
\begin{figure}[b!]
    \centering
    \includegraphics{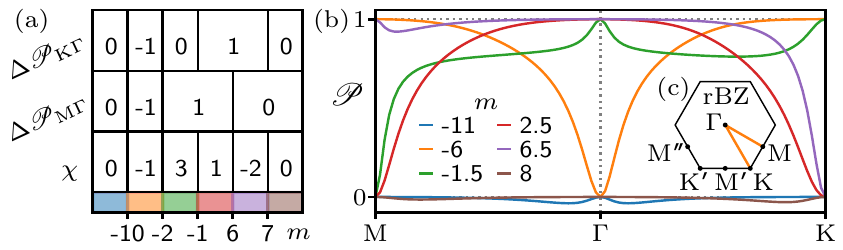}
        \caption{(a) The polarization and Hopf invariants as a function of $m{\in}\mathbb{R}$ [cf.\ Eq.~(\ref{eq:hopfc6ham})]; colors distinguish distinct phases. (b) For each phase, the polarization along $\textrm{M}\Gamma \textrm{K}$ is plotted for a representative value of $m$. (c) Rotation-invariant points in reduced Brillouin zone.} \label{fig:RTP}
\end{figure}

\emph{Minimal model.---} To exemplify a non-trivial RTP, we consider a two-band, tight-binding model with six-fold ($C_6$) rotational symmetry. On each site of a triangular lattice, we situate an $s$ and a $p_+{=}p_x{+}ip_y$ (spinless) orbital, which transform under $C_6$ with angular momenta {$\ell_v{=}0$ and $\ell_c{=}1$}, respectively. The Hamiltonian has the form
\begin{eqnarray}
\begin{aligned}
{H}(\bk) \eq [z^{\dagger}(\bk)\bm{\sigma}z(\bk)]\cdot\bm{\sigma},\;\;\;\;z(\bk)=(z_1,z_2)^T \\ 
z_1(\bk) \eq \sum_{a=1}^{6}e^{-i\pi a/3 } \exp[i\bm{t}(a)\cdot\bk_\perp], \\ 
z_2(\bk) \eq \sin k_z + i\left(\sum_{a=1}^{6}\exp[i\bm{t}(a)\cdot\bk_\perp] + 4\cos k_z + m\right),
\end{aligned} \label{eq:hopfc6ham}
\end{eqnarray}
with  $\bm{t}(a){=}[\cos(\pi a/3), \sin(\pi a/3)]$, $m$ a tuning parameter for topological-phase transitions, and $\bm{\sigma}$ the vector of Pauli matrices with $\langle\sigma_z\rangle{=}1$ (resp.\ $-1$) corresponding to the $s$ (resp.\ $p_+$) orbital. For generic $m{\in}\mathbb{R}$, an energy gap exists throughout the BZ, and the conduction (resp. valence) eigenvector is a periodic-in-BZ, analytic function  $u^c(\bk){=}z(\bk)/||z(\bk)||$  (resp.~$u^v(\bk){=}i\sigma_y u^c(\bk)^*$) satisfying the symmetry condition $U_{6} u^{v/c}(\bk){=}\exp(i2\pi \ell_{v/c}/6)u^{v/c}(R_{6}\bk)$, with $U_6$ (resp.\ $R_6$) the pseudospinor (defining) representation of $C_6$. 
Consequently, the mutually-disjoint condition is satisfied with $u^c$ (resp. $u^v$) being $p_+$-like (resp.\ $s$-like) along all rotation-invariant lines, and $u^{c/v}$  Fourier transform to symmetric, exponentially-decaying\cite{cloizeaux-analyt-wannier} WFs [Fig.~\ref{fig:wannier}(a,b)].  Applying our previous argument for the integer-quantization of $\Delta\mathscr{P}_{\bkp'\bkp''}$, we find that the polarization at all $C_2$-invariant points ($\Gamma,\textrm{M},\textrm{M}',\textrm{M}''$) and $C_3$-invariant points ($\Gamma,\textrm{K},\textrm{K}'$) in the rBZ [cf.\ Fig.~\ref{fig:RTP}(c)] are identical modulo integer. The six-fold symmetry implies there are two independent polarization differences $\Delta\mathscr{P}_{\textrm{K}\Gamma}$ and $\Delta\mathscr{P}_{\tm\Gamma}$.

For large $\abs{m}$, the Hamiltonian reduces to a $\bk$-independent diagonal form $H(\bk){\approx}{-}m^2 \sigma_z$, implying that the $s$-type valence (and also the $p_+$-type conduction) band is unicellular. This is consistent with $\mathscr{P}(\bkp)$ being continuously deformable to a flat sheet for representative values $m{=}{-}11$ and $m{=}8$, as illustrated by the blue resp.~brown line in Fig.~\ref{fig:RTP}(b). Increasing $m$ from $-11$ to $-10$, the bulk gap closes at the Brillouin-zone center; the resultant effective-mass Hamiltonian has the form in  \q{eq:hopfc6ham} with $z_1{=}3(k_y{+}ik_x)$ and $z_2{=}k_z{+}i(10{+}m)$, which identifies the quadratic band-touching point as a \textit{dipole} source of Berry curvature~\cite{AA_teleportation} with dipole moment parallel to the rotation-invariant $\bk$-line. This dipole intermediates~\cite{AA_teleportation} a valence-to-conduction transfer of a  $2\pi$ quantum of the Berry-Zak phase ($\phi_Z$)  -- defined for the parallel transport of Bloch functions along said $\bk$-line. Since $\phi_Z/2\pi{\equiv}_1 \mathscr{P}(\Gamma)$ according to the geometric theory of polarization~\cite{zak_berryphase,kingsmith_polarization}, with  $\equiv_j$ meaning ``equal (mod $j$)'', there is correspondingly a discontinuous, unit-decrease of $\Delta\mathscr{P}_{\textrm{M}\Gamma}$ and $\Delta\mathscr{P}_{\textrm{K}\Gamma}$ when the gap reopens for $\delta m{:}{=}m{+}10{\gtrsim}0$ [orange line in Fig.~\ref{fig:RTP}(b)]. This further manifests as a ``$(\delta m)^2\sgn[\delta m]$''-type non-analyticity in the total polarization $\overline{\mathscr{P}}{=}\int d^2 k_{\perp} \mathscr{P}(\bk_{\perp})/\textrm{Area}(\textrm{rBZ})$ [cf.~\fig{fig:wannier}(c--d)]~\cite{supp}. Further gap closings (at $m{=}{-}2,{-}1,6,7$) result in Berry dipoles at other high-symmetry wavevectors, with the resultant phase diagram and RTP's summarized in Fig.~\ref{fig:RTP}(a,b).    

\emph{Stability of RTP.---} Equation~(\ref{eq:hopfc6ham}) represents a minimal model of an  RTP with the smallest dimension for the matrix $H(\bk)$. Models of arbitrarily large matrix dimensions can be constructed from our minimal model by adding unicellular bands to either  conduction or valence subspace, assuming their symmetry representations maintain the mutually-disjoint condition -- this preserves the integer-valued quantization of  $\Delta\mathscr{P}_{\textrm{M}\Gamma}$ and $\Delta\mathscr{P}_{\textrm{K}\Gamma}$, hence also the RTP. In contrast (as numerically verified in the Supplemental Material (SM)~\cite{supp}), the quantization is lost upon addition of unicellular conduction bands that nullify the mutually-disjoint condition, thus manifesting the RTP is a symmetry-protected delicate invariant. 

\emph{Multicellularity with only translational symmetry.---} Which of our conclusions survive when rotational symmetry is relaxed? While the RTP generically destabilizes, we show that multicellularity persists -- at least for the minimal model and any continuous deformation thereof that preserves the bulk energy gap and the bulk \textit{translational} symmetry; any other symmetry can be relaxed. We appeal to a special feature of Pauli-matrix Hamiltonians with a spectral gap at each three-momentum $\bk$; namely, that even with a trivial first Chern class, $H(\bk)$ has an integer-valued classification given by the Hopf invariant $\chi$~\cite{Hopf:1931,pontrjagin_classification,kennedy_hopfchern,unal_floquethopf,Hopfinsulator_Moore} which is equivalent to a Brillouin-zone ($\textrm{BZ}$) integral of the Abelian Chern-Simons three-form~\cite{Hopfinsulator_Moore,wilczekzee_linkingnumbers} \e{\chi=-\f1{4\pi^2}\int_\textrm{BZ} \bA\cdot \left(\nabla{\times}\bA\right)  \,d^3k,\la{definechi}} with  $\bA(\bk){=}{\braket{u}{i\nabk u}}$ the Berry connection of an energy-nondegenerate band~\cite{berry_quantalphase}. Since $\chi$ is integer-quantized only for Pauli-matrix Hamiltonians, it is manifestly a delicate topological invariant distinct from RTP. That our minimal model for $m{\in}[{-}10,{-}2]$ has $\chi{=}{-}1$ is a consequence of a single Berry dipole intermediating a unit change in~$\chi$ at $\Gamma$ at $m{=}{-}10$~\cite{AA_teleportation}.

That $\chi{\neq}0$ implies multicellularity is now proven by contradiction. Assume that the valence-band WF is localizable to one unit cell, i.e., $W_{\bR}^{v}{=}\delta_{\bR,\bze}\kappa_v$, with $\kappa_v$ a pseudo-spinor wave function that corresponds to a single point on the Bloch sphere $S^2$. The Fourier transform of $W_{\bR}^v$ is then $\bk$-independent, namely $u_{v}(\bk){=}\kappa_v$. It is an eigenvector of a Hamiltonian that represents the trivial, constant map from the BZ to $S^2$, in contradiction with the assumed non-trivial Hopf invariant. 

\emph{Hopf-RTP correspondence.---} We have shown that \textit{both} the Hopf invariant and RTP imply multicellularity. In fact, by a straightforward application of Whitehead's formulation of the Hopf invariant~\cite{Whitehead:1947}, we find~\cite{supp} that the Hopf invariant and RTP are related to each other as [cf.~Fig.~\ref{fig:RTP}(a)]
\begin{equation}
\chi\;\equiv_6    3\Delta\mathscr{P}_{\textrm{M}\Gamma} - 2\Delta\mathscr{P}_{\textrm{K}\Gamma},
    \label{eq:RTP-hopf-C6}
\end{equation}
for any $C_6$-symmetric, Pauli-matrix Hamiltonian having trivial Chern class and satisfying the uniaxial and mutually-disjoint conditions with $\ell_v{=}0,\ell_c{=}1$.

\emph{Bulk-boundary correspondence.---} We have established the RTP and Hopf invariant as bulk delicate invariants leading to bulk multicellularity, but what does bulk multicellularity imply in the presence of a rotation-invariant surface termination? We answer with the following \textit{obstruction principle}: there does not exist a symmetric, 2D tight-binding description (of a single surface facet) where all WFs are centered on the same rotational axes as the bulk WFs. Alternatively stated, on a half-infinite slab, the \textit{entire} Hilbert space of  states -- filled and unfilled, bulk-extended and surface-localized -- cannot be spanned by (uniaxially-symmetric, exponentially-localized) WFs whose positional centers coincide with the WFs obtained under periodic boundary conditions. (In contrast, for the `boundary-obstructed' topological phase studied in \ocite{khalaf2021}, a Wannier obstruction exists for the filled subspace but not for the entire Hilbert space.)

\begin{figure}[b!]
    \centering
    \includegraphics{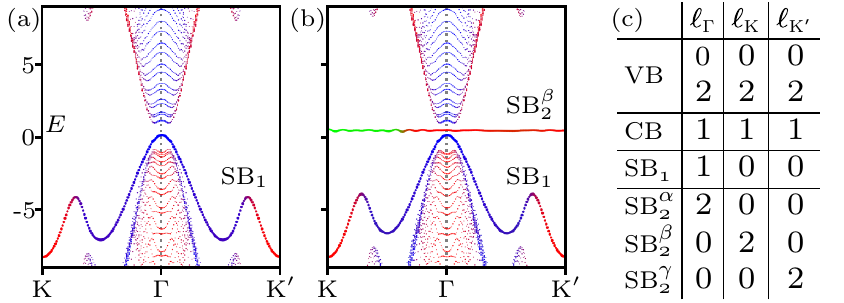}
    \caption{Spectrum of (a) Hopf-insulating and (b) RTP-insulating slab with one resp.~two detached surface bands. Red/blue/green coloring represents contribution to the bands from basis orbitals with $\ell{=}0/1/2$.     (c) For various bands discussed in the main text, $\ell_\Gamma{,}\ell_\tk{,}\ell_\tkpr$ denote the mod-three angular momenta at $C_3$-invariant wavevectors.} \label{fig:bbc}
\end{figure}

A stronger form of our principle is realized by the half-infinite, Hopf-insulating slab (with or without rotational symmetry), namely that its Hilbert space does not even have an exponentially-localized WF representation, because it is characterized \cite{Brouder_explocWFs} by a nonvanishing first Chern number -- a stable, $K$-theoretic invariant~\cite{kitaev_periodictable}. This follows from the equality~\cite{AA_teleportation} of the bulk invariant $\chi$ and the \textit{faceted Chern number} $\mathscr{C}_f$ -- defined as the net Chern number of all surface-localized bands, \textit{independent} of filling. The reason for this bulk-boundary correspondence is that bulk bands (characterized by a nontrivial Chern-Simons three-form of the Berry connection [cf.\ \q{definechi}]) result in a surface anomalous Hall conductance (SAHC), according to the geometric theory of the magnetoelectric polarizability\cite{essin_magnetoelectric, qi_topologicalfieldtheory, essin_magnetoelectric_erratum, essin_orbitalmagnetoelectric_bandinsulators, malashevich_magnetoelectric, vanderbilt_book_berryelectronicstructure}; since the net SAHC of the entire Hilbert space must vanish, this necessitates the existence of surface bands which contribute a cancelling SAHC~\cite{AA_teleportation}. Figure~\ref{fig:bbc}(a) illustrates the topologically nontrivial surface-localized band with  Chern number $\mathscr{C}_f{=}{-}1$ for our minimal model  ($m{=}{-}6$, $\chi{=}{-}1$); we emphasize that band(s) with the counter-balancing Chern number $\mathscr{C}_f'{=}{+}1$ do not exist  in the entire Hilbert space (of filled and unfilled states) on a half-infinite geometry. 
 
Next we demonstrate with an example that the Hilbert space of a half-infinite RTP insulator either has no 2D tight-binding description (owing to stable or fragile topology), or has a 2D tight-binding description with displaced Wannier centers. To  model an insulator that is not a Hopf insulator and yet has a nontrivial RTP, we enlarge the Hilbert space of our minimal model ($m{=}{-}6$) by adding a unicellular valence band whose representative WF has angular momentum  $\ell{=}2$. To simplify the discussion, we restrict ourselves to the $P3$ space group by including $C_3$-symmetric (and $C_2$-asymmetric) Hamiltonian matrix elements. By construction, the mutually-disjoint condition is satisfied for representations of $C_3$, thus the polarization difference $\Delta\mathscr{P}_{\textrm{K}\Gamma}{=}{-}1$ remains quantized, but quantization  no longer holds for $\Delta\mathscr{P}_{\textrm{M}\Gamma}$.

For the bulk valence ($\textrm{VB}$) and conduction bands ($\textrm{CB}$), the symmetry representations at $C_3$-invariant wavevectors are presented in the first three rows of Fig.~\ref{fig:bbc}(c). The fourth row of Fig.~\ref{fig:bbc}(c) gives the symmetry representations of the nontrivial surface band $\textrm{SB}_1$ [cf.~Fig.~\ref{fig:bbc}(b)], which is topologically equivalent to the nontrivial surface band of the minimal model in Fig.~\ref{fig:bbc}(a). Observe that the representations of $\textrm{SB}_1$ are identical to those of $\textrm{VB}$ \textit{except} at $\Gamma$, where $\textrm{SB}_1$ has the same representation as $\textrm{CB}$. This may be rationalized by a thought experiment of imposing a surface termination on the bulk $s$-dominated WF; because of its nontrivial polarization [cf. \fig{fig:wannier}(c)], such termination generates energetically-unfavorable dangling bonds; to remove these bonds, the surface WF hybridizes with $p_+$-type orbitals.

We are ready to diagnose the advertised obstruction: having Chern number $\mathscr{C}_{f}{=}{-}1$, $\textrm{SB}_1$ has no exponentially-localized WF representation. To attain such a representation, one must sum the surface band with another band over the $\textrm{rBZ}$ having the opposite Chern number. Indeed, by modification of the surface Hamiltonian, one may always localize a second surface band $\textrm{SB}_2$ by detaching it (i.e., `peeling it off') from either $\textrm{VB}$ or $\textrm{CB}$. If detached from $\textrm{CB}$, $\textrm{SB}_2$ combines bulk symmetry representations from the conduction  subspace [third row in Fig.~\ref{fig:bbc}(c)]. For a $C_3$-symmetric band with Chern number $\mathscr{C}$, the product of  $C_3$ eigenvalues at $\{\Gamma{,}\textrm{K}{,}\textrm{K}'\}$ gives $e^{-i2\pi \mathscr{C}/3}$~\cite{Chen_bulktopologicalinvariants}. It follows that any detachment from $\textrm{CB}$ necessarily has $\mathscr{C}{\equiv_3} 0$, and cannot nullify the unit Chern number of $\textrm{SB}_1$. Instead, if we apply the same rule to detachments from $\textrm{VB}$, we find three possible symmetry representations for $\textrm{SB}_2$ with $\mathscr{C}{=}{+}1$, which we denote by $\textrm{SB}^{\alpha,\beta,\gamma}_2$ in Fig.~\ref{fig:bbc}(c) and discuss in order.

Though a WF representation exists for the composite band $\textrm{SB}_1{\oplus} \textrm{SB}_2^\alpha$, these WFs cannot individually be $C_3$-symmetric on any of the  $C_3$-invariant Wyckoff positions: $\{1a,1b,1c\}$. Indeed, the symmetry representations of $\textrm{SB}_1{\oplus} \textrm{SB}_2^{\alpha}$ are incompatible with a band representation of space group $P3$, which is deducible by comparison with symmetry-representation tables in the Bilbao crystallographic server~\cite{elcoro_EBRinBilbao}. The obstruction to $C_3$-symmetric WFs is \emph{fragile}, in the sense that a trivial band $\textrm{TB}$ exists (though not necessarily in the present Hilbert space),  such that $\textrm{SB}_1{\oplus} \textrm{SB}_2^{\alpha}{\oplus}\textrm{TB}$ is not obstructed. 

In contrast, by comparing the symmetry representations of $\textrm{SB}_1{\oplus}\textrm{SB}_2^{\beta}$ with the Bilbao tables, we deduce that $\textrm{SB}_1{\oplus} \textrm{SB}_2^\beta$ is a band representation with representative WFs of angular momentum $\ell{=}1$ and $\ell{=}0$, centered on the $1c$ and $1a$ Wyckoff positions, respectively~\cite{supp} [Fig.~\ref{fig:wannier}(e,f)]; $\textrm{SB}_1{\oplus} \textrm{SB}_2^\gamma$ is likewise band-representable with $\ell{=}1$ and $\ell{=}0$, centered on $1b$ and $1a$, respectively. Indeed, no matter how many bands are detached from $\textrm{VB}$  and added to $\textrm{SB}_{1}$, the resultant, composite band cannot have a tight-binding description with all Wannier centers on the $1a$ Wyckoff position of the bulk WFs. (In the language of Topological Quantum Chemistry,\cite{TQC} the surface WFs realize an `obstructed atomic limit', while this is not true for the bulk WFs.\cite{supp}) Assuming the contrary, the set of $C_3$ eigenvalues of the composite band must be identical at $\Gamma,\textrm{K}$ and  $\textrm{K}'$.\cite{TBO_JHAA} But this cannot be satisfied, because $\textrm{SB}_1$ contributes one $C_3$ eigenvalue (${=}e^{i2\pi/3}$) at $\Gamma$ which can never have an equal counterpart at $\textrm{K}$ and $\textrm{K}'$.

\emph{Conclusion.---} The multicellular landscape, as enriched by crystalline symmetries, promises to be fertile ground for TIs that would naively be missed and deemed trivial. We have introduced two (not necessarily disjoint) classes of multicellular, Wannierizable TIs: rotation-invariant insulators with a returning Thouless pump (RTP), and Hopf insulators. For both classes, we have shown that bulk multicellularity (a) is a delicate topological invariant, and (b) implies that the Hilbert space (on a half-infinite slab) cannot be Wannierized with WF centers identical to those of the bulk WFs. Whether (a--b) extend to \textit{all} multicellular TIs is presently unanswered. Whether all delicate topological invariants are accompanied by bulk multicellularity is also unknown. 

Our formulation of the RTP in terms of the Berry-Zak phase allows for a high-throughput search for materials candidate. We have identified over forty hexagonal magnetic space groups that allow a symmetry-protected RTP, which we tabulated in Sec.~IX of SM~\cite{supp}. After selecting materials in these space groups whose low-energy bands satisfy the mutually-disjoint symmetry condition, one would compute the Berry-Zak phase by standard first-principles techniques \cite{z2pack}.

The multicellular Hopf insulator is already known to manifest  higher-order topology, quantized surface magnetism~\cite{penghaoAA_quantizedmagnetism}, and quantized magneto-electric polarizability~\cite{AA_teleportation}; it would be interesting to investigate if these properties extend to other multicellular/delicate topological insulators. Beyond band theory, we expect  multicellularity to add a new chapter to the interplay between non-unicellular WFs, generalized Hubbard models and exotic correlated phases~\cite{hofmann_berg_flatbandSC,kang_vafek_tblg,peri_bernevig_flatbandSC}.

\emph{Acknowledgments.---} 
We thank A.~Bouhon for alerting us to Whitehead's formulation of the Hopf invariant, and acknowledge a stimulating discussion with B. A.~Bernevig about the obstructed atomic limit. Zhida Song helped to clarify a question on symmetry indicators. A.~N. was supported by the Swiss National Science Foundation (SNSF) grant No.~176877, and by Forschungskredit of the University of Z\"{u}rich, grant No.~FK-20-098. T.~N. acknowledges support from the European Research Council (ERC) under the European Union’s Horizon 2020 research and innovation programm (ERC-StG-Neupert-757867-PARATOP) and from NCCR MARVEL funded by the SNSF. 
T.~B. was supported by the SNSF Ambizione grant No.~185806. A.~A. was supported by the Gordon and Betty Moore Foundation EPiQS Initiative through Grant No. GBMF 4305 and GBMF 8691 at the University of Illinois.

\let\oldaddcontentsline\addcontentsline     
\renewcommand{\addcontentsline}[3]{}        

\let\addcontentsline\oldaddcontentsline

\clearpage

\renewcommand{\theequation}{S\arabic{equation}}
\renewcommand{\thefigure}{S\arabic{figure}}
\renewcommand{\thetable}{S\arabic{table}}
\renewcommand{\thepage}{\roman{page}}
\setcounter{page}{1}
\setcounter{equation}{0}
\stepcounter{myfigure}
\setcounter{affil}{0}
\renewcommand\thesection{\Alph{section}}
\renewcommand\thesubsection{\arabic{subsection}}


\newpage
\onecolumngrid
\begin{center}
    {\large \textbf{Supplemental Material to: Multicellularity of delicate topological insulators}}
    \\
    \vspace{\baselineskip}
    Aleksandra Nelson,\textsuperscript{1} Titus Neupert,\textsuperscript{1} Tom\'{a}\v{s} Bzdu\v{s}ek,\textsuperscript{2, 1} and A. Alexandradinata\textsuperscript{3}
    \\
    {\small \vspace{\baselineskip}
    \textsuperscript{1}\textit{Department of Physics, University of Zurich, Winterthurerstrasse 190, 8057 Zurich, Switzerland}
    \\
    \textsuperscript{2}\textit{Condensed Matter Theory Group, Paul Scherrer Institute, 5232 Villigen PSI, Switzerland}
    \\
    \textsuperscript{3}\textit{Department of Physics, University of Illinois at Urbana-Champaign, Urbana, Illinois 61801-2918, USA}
    \\
    (Dated: \today)}
\end{center}
{\tableofcontents \par}

\section{\label{sec:OAL}Multicellularity vs the obstructed atomic limit}

Let us comment on the intersection between multicellular topological insulators and obstructed atomic insulators, and show that neither of these notions necessarily implies the other. 

\subsection{An elaboration on the definition of unicellularity}\la{sec:elaboration}

At the onset, it is worth elaborating on the definition of a unicellular band, which was briefly stated in the main text as there existing a set of symmetry-respecting, exponentially-localized Wannier functions that span the band's Hilbert space, and each Wannier function can be confined to a single primitive unit cell by a continuous, adiabatic deformation of the Hamiltonian. 

We consider a Hilbert space that is spanned by an orthonormal Wannier basis $\{\varphi_{\bR,\alpha}\}_{\bR\in \textrm{BL},\alpha=1\ldots \mathcal{V}+\mathcal{C}}$ where $\textrm{BL}$ denotes the Bravais lattice and $\mathcal{V}$ ($\mathcal{C}$) will play the role of the number of valence (occupied) bands of the studied model. Each basis vector (or basis `orbital') is localized to a single lattice site,  which we can formalize by specifying how the discrete position operator acts:
\e{\hat{\br}\ket{\varphi_{\bR,\alpha}}=(\bR+\bw_{\alpha})\ket{\varphi_{\bR,\alpha}},\la{formalize}}
where $\bs{R}$ are Bravais lattice vectors; the physical positions of Wannier orbitals within one unit cell need not be coincident and are distinguished by  $\bw_{\alpha}$.

For a given Hamiltonian acting on this Hilbert space a collection of $V$ occupied bands, labelled by $\nu\in\{1,2,\ldots,V\}$, has the following Wannier representation
\begin{equation}
\ket{W^\nu_{\bs{R}}} = \sum_{\alpha,\br} c^\nu_{\br,\alpha} \ket{\varphi_{\bs{R}-\br,\alpha}}.\label{eqn:Wanier-general}
\end{equation}
The coefficients $c_{\br,\alpha}^\nu$ are complex amplitudes (with magnitudes squared normalized to $1$ after summing over orbitals $\alpha$ and Bravais vectors $\br$).

We call a collection of Wannier functions $\left\{\ket{W_{\bs{R}}^\nu}\right\}_{\nu=1}^V$ \emph{unicellular} if only orbitals with $\br=\bs{0}$ contribute to the sum on the right side of Eq.~(\ref{eqn:Wanier-general}) for each $\nu$, i.e., they acquire the form
\begin{equation}
\ket{W^\nu_{\bR}} = \sum_{\alpha} c^\nu_{\alpha} \ket{\varphi_{\bs{R},\alpha}}.\label{eqn:Wanier-strict-uni}
\end{equation}
\\

A few clarifying remarks are in order:\\

\noi{a} Recall that a primitive unit cell is a finite region of space that, when translated by the Bravais-lattice vectors, covers all space ($\mathbb{R}^d$) without overlapping. While the volume of this finite region is uniquely defined by the Bravais lattice, its boundary is not. In the definition of unicellularity, we assume that we can find a unit cell such that \emph{all} $V$ representative, valence-band Wannier functions are strictly localized within it. [A representative set of valence-band Wannier functions are a  minimal set of Wannier functions which generate an infinite set of Wannier functions (under Bravais-lattice translations) that span the valence band. ]\\

\noi{b} One could adopt a generalized (i.e., less restrictive) definition of unicellularity, namely that Wannier functions can be constructed such that each one of them can be confined to an appropriately chosen unit cell, while not requiring these choices of unit cells to be equal to each other. 
We emphasize that (unless explicitly stated otherwise, see e.g.~the footnote in Sec.~\ref{RTP-multicel}) we adopt the ``strict'' definition of unicellularity from (a) throughout the main text and supplemental material.
\\

\noi{c} Throughout this work, we deal with Wannier functions in the tight-binding formalism. Tight-binding Wannier functions are defined over a set of discrete spatial points (`sites'), rather than continuous space. By `symmetry-respecting, exponentially-localized Wannier functions', we mean precisely that the band (spanned by said Wannier functions) is a band representation~\cite{Zak_bandrepresentations}, namely it is a representation of a space group $G$ induced from a representation of site stabilizer $G_{\bw}$ (defined as the subgroup of $G$ that preserves the spatial coordinate $\bw$). The sum of two band representations (in the sense of a Whitney sum of the two corresponding vector bundles) is also defined 
to be a band representation.\\

\noi{d} In the following we need to convert Bloch eigenstates $\ket{\psi_{\bk}^\nu}$ to the Wannier functions $\ket{W_{\bR}^\nu}$ and vice versa. Throughout the whole text we use the following convention
\e{
    \ket{W_{\bR}^\nu} \eq \frac{1}{\textrm{Vol}(\textrm{BZ})}\int d\bk e^{-i\bk\cdot\bR}\ket{\psi_k^\nu}, \lin
    \ket{\psi_k^\nu} \eq \sum_{\bR} e^{i\bk\cdot\bR}\ket{W_{\bR}^\nu}
    \label{eq:bloch-to-wannier}
}
where $\textrm{Vol}(\textrm{BZ})$ is the volume of the Brillouin zone. The same convention can be applied to the wavefunctions in tight-binding formalism in which given the Wannier basis $\ket{\varphi_{\bR,\alpha}}$ the Wannier functions are defined by the vector of coefficients $c^\nu_{\br}$ from Eq.~\eqref{eqn:Wanier-general}. To define the Bloch functions we first specify a basis of Bloch states, which are obtained by Fourier transforming the Wannier basis orbitals
\e{
\ket{\Phi_{\bk,\alpha}}=\sum_{\bR}e^{i\bk\cdot(\bR+\bw_{\alpha}) }\ket{\varphi_{\bR,\alpha}}, 
\label{eq:bloch_basis}
}
where we included spacial position of the basis orbitals $\bw_\alpha$.
Then the Bloch states are defined by the Bloch vectors $u^\nu(\bk)$
\e{
\ket{\psi_k^\nu} \eq \sum_\alpha u^\nu_\alpha(\bk)\ket{\Phi_{\bk,\alpha}} 
\label{eq:tight-binding-vectors}
}
The vectors of coefficients ${c}_{\br}^\nu=\{c_{\br,\alpha}^\nu\}_{\alpha=1,\ldots,\mathcal{V}+\mathcal{C}}$ and ${u}^\nu(\bk)=\{u_\alpha^\nu(\bs{k})\}_{\alpha=1,\ldots,\mathcal{V}+\mathcal{C}}$ are related by the  transformation
\begin{equation}
    {c}^\nu_{\br} = \frac{1}{\textrm{Vol}(\textrm{BZ})}\int d\bk e^{-i\bk\cdot(\br-\hat{\bw})} {u}^\nu(\bk)
    \label{eq:bloch-to-wannier-tb}
\end{equation}
where $\hat{\bw}$ is a diagonal  matrix having diagonal elements equal to $\{\bw_{\alpha}\}_{\alpha=1\ldots \mathcal{V}+\mathcal{C}}$.\\

\noi{e} With regard to `continuous, adiabatic deformation of the Hamiltonian', in addition to deformations of the matrix elements for a fixed tight-binding basis, we also allow for deformations of the tight-binding Hilbert space that maintains  a pre-specified crystallographic space group $G$. \\

\noindent To clarify what (e) means, note that the basis 
$\{\varphi_{\bR,\alpha}\}_{\bR\in \textrm{BL},\alpha=1\ldots \mathcal{V}+\mathcal{C}}$ is itself a band representation of $G$.
Matrix elements of a tight-binding Hamiltonian $H(\bk)$ in the momentum representation are defined with respect to a basis of Bloch states [Eq.~\eqref{eq:bloch_basis}].
Because the Bloch states are not generally periodic over the Brillouin zone, likewise
\e{ H(\bk+\bG) = e^{-i\bG\cdot \hat{\bw}}H(\bk)e^{i\bG\cdot \hat{\bw}},\as \bG \in \textrm{RL}\label{eqn:the-w-hat}}
with $\bG$ any vector of the reciprocal lattice (\textrm{RL}).
We allow to deform the tight-binding Hilbert space in two ways. \\

\noi{e-i} Firstly, consider the set of all basis vectors lying on the same position $\bR+\bw_{\alpha}$, which span a finite-dimensional Hilbert space $\calh_{\bR+\bw_{\alpha}}$; because the tight-binding Hilbert space is a band representation of $G$, $\calh_{\bR+\bw_{\alpha}}$ must form a representation of the site stabilizer $G_{\bR+\bw_{\alpha}}$. We allow for any unitary transformation within $\calh_{\bR+\bw_{\alpha}}$.\\

\noi{e-ii} Secondly, we allow to continuously displace the basis Wannier centers as $\hat{\bw} \ri \hat{\bw}+\delta \hat{\bw}$. Its effect on the momentum-dependent Hamiltonian is  a unitary transformation:
\e{ H(\bk)\ri e^{-i\bk\cdot \delta \hat{\bw}}H(\bk)e^{i\bk\cdot \delta \hat{\bw}}.}
Being unitary, such a deformation will not affect energies, and is automatically adiabatic. We only allow displacements that preserve the band-representability of the tight-binding Hilbert space. \\

\noindent \textit{Example of a deformation of the type (e-ii).---} Let us consider the uniaxial tight-binding models introduced in the main text, where all of $\{\bR+\bw_{\alpha}\}$ lie on rotation-invariant axes. In the minimal model of Eq. (1) in the main text,  $\bw_{\alpha}=0$ for all $\alpha$, hence $H(\bk)$ is periodic, and the polarization $P(\bkp)$ at $C_n$-invariant wavevectors are quantized to integers, as  verifiable in Fig.\ 1(c) in the main text. An allowed deformation that maintains rotational symmetry is to move a basis orbital along a rotational axis; to maintain translational symmetry, all basis orbitals related by a Bravais-lattice translation must simultaneously be moved. It would follow that  $P(\bkp)\equiv_1 \sum_{\alpha} [\bw_{\alpha}]_z$ is not generally integer-valued, with $[\bw_{\alpha}]_z$ here referring to the coordinate on the rotational axis. Yet, differences in $P(\bkp)$ over distinct $C_n$-invariant wavevectors remains quantized, and hence the returning Thouless pump (RTP) remains well-defined. 

\subsection{Multicellular topological insulators are not necessarily obstructed atomic insulators}

According to the theory of topological quantum chemistry~\cite{cano_buildingblocksTQC}, the obstructed atomic insulator is a band representation whose valence band is spanned by a set of symmetric exponentially-localized Wannier functions $\{W^{v}_{j,\bR}\}_{\bR\in \textrm{BL},j=1\ldots \mathcal{V}}$, whose corresponding 
Wannier centers  do not coincide with the `atomic positions' (to be clarified below); moreover, owing to certain crystallographic point-group symmetries that fix each Wannier center to a high-symmetry Wyckoff position, the Wannier center cannot be smoothly deformed to the atomic positions which are assumed to lie on a distinct Wyckoff position. From the perspective of tight-binding models, the `atomic positions' are naturally identified with the positions of basis vectors, as given in \q{formalize}.

According to this definition, our proposed $C_6$-symmetric, RTP insulator[cf.\ ~\eqref{eq:hopfc6ham}]  is not an obstructed atomic insulator, as its valence  Wannier functions are centered at the same Wyckoff position $1a$ (with site stabilizer $C_6$) as the basis `atomic' orbitals. A second example is the Hopf insulator without point-group symmetry. {Because the Wannier center (of the valence subspace) is movable without symmetry restriction, it is (trivially) not an obstructed atomic insulator; yet, the Hopf insulator remains multicellular, based on an argument presented in the main text.}

\begin{figure}
    \centering
    \includegraphics{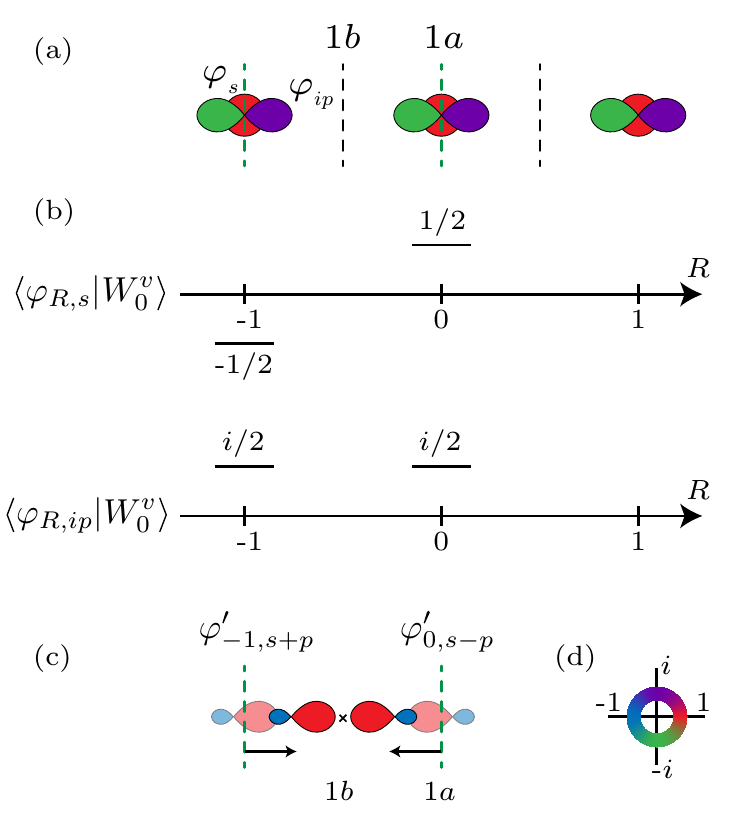}
    \caption{(a) SSH-type model with inversion-even $\varphi_s$ and inversion-odd $\varphi_{ip}$ tight-binding basis orbitals localized at $1a$ Wyckoff positions. The color indicates the amplitude of the orbital's wave function according to the color scheme displayed in panel (d). The primitive unit cells centered at $1a$ resp.~$1b$ Wyckoff positions are bounded by black resp.~green dashed lines. (b) The valence Wannier function is a linear combination of basis orbitals at neighboring sites. The coefficients $\braket{\varphi_{R,\alpha}}{W^v_0}$ are presented as a function of the Bravais vector $R$. (c) The hybridized orbitals $\varphi^\prime_{-1, s+p}$ and $\varphi^\prime_{0, s-p}$ that contribute to the valence Wannier function are formed from basis orbitals with coefficients denoted by colors from the color wheel (d). Their centers can be shifted to position $1b$ while preserving the inversion symmetry.
    }
    \label{fig:SSH}
\end{figure}

\subsection{Obstructed atomic insulators are not necessarily multicellular}

The title of this subsection is exemplified by an inversion-symmetric Su-Schrieffer-Heeger- (SSH-)~\cite{su1979} type model, whose tight-binding basis is given in each unit cell by an inversion-even $s$ and an inversion-odd $ip$ orbital localized to the same inversion-invariant Wyckoff position $1a$ [cf.~Fig.~\ref{fig:SSH}(a)]. In this basis, the matrix representation of inversion (about center $1a$) is $\sigma_z$, while time-reversal symmetry is represented by $\sigma_z K$ with $K$ being the complex conjugation. A representative tight-binding Hamiltonian is given by
\begin{equation}
    H(k)=\cos k\sigma_z+\sin k\sigma_x, \as \sigma_z H(k)\sigma_z=H(-k).
\end{equation}
To obtain exponentially-localized Wannier functions for the energy bands, we need to find eigenvectors of $H(k)$ that are smooth and periodic over the Brillouin zone. In spite of our model Hamiltonian $H(\bk)$ being real, the requirements of smoothness and periodicity can only be fulfilled by complex-valued eigenvectors, because both valence and conduction bands of our model Hamiltonian have a nontrivial first Stiefel-Whitney class (i.e., they carry $\pi$ Berry phase)~\cite{nogo_AAJH,ahn_stiefelwhitney}. One choice of smooth, periodic eigenvectors is
\begin{equation}
    u^v(k)=-ie^{ik/2}\begin{pmatrix}
    \sin k/2 \\ -\cos k/2 
    \end{pmatrix}, \quad
    u^c(k)=-ie^{ik/2}\begin{pmatrix}
    \cos k/2 \\ \sin k/2 
    \end{pmatrix}.
\end{equation}
for the valence and conduction subspace respectively. 
The corresponding valence Wannier function is centered at the $1b$ `mid-bond' Wyckoff position and is formed by an inversion-odd linear combination of atomic orbitals from two nearest-neighbor $1a$ sites [Fig.~\ref{fig:SSH}(b)] as can be derived by combining Eqs.~(\ref{eqn:Wanier-general},\ref{eq:bloch-to-wannier-tb}): 
\begin{equation}
    \braket{\varphi_{R,\alpha}}{W^v_0}=\int \f{dk}{2\pi} e^{ik R}u^v_\alpha(k)= (\delta_{R,0}\delta_{\alpha,s}-\delta_{R,-1}\delta_{\alpha,s}+i\delta_{R,0}\delta_{\alpha,ip}+i\delta_{R,-1}\delta_{\alpha,ip})/2,
\end{equation} 
To express this in simpler terms, we can define a new $sp$-hybridized basis $\{\varphi'_{R, s-p},\varphi'_{R,s+p}\}_{R\in \mathbb{Z}}$ 
such that each new basis vector is a  linear combination of the old basis vectors (on the same site):
\begin{subequations}
\begin{eqnarray}
    \ket{\varphi'_{R,s-p}} &=& \frac{1}{\sqrt{2}}\left(\ket{\varphi_{R, s}}+i\ket{\varphi_{R, ip}}\right), \\
    \ket{\varphi'_{R,s+p}} &=& \frac{1}{\sqrt{2}}\left(\ket{\varphi_{R, s}}-i\ket{\varphi_{R, ip}}\right).
\end{eqnarray}\la{basischange}
\end{subequations}
such that the valence Wannier function is simply the sum:
\e{ W^v_0=\f1{\sqrt{2}}\left(\varphi'_{0, s-p}-\varphi'_{-1,s+p}\right) \la{RHS}}
as illustrated in Fig.~\ref{fig:SSH}(c).
[Note \q{basischange} is a basis change of the type (e-i) discussed in \s{sec:elaboration}.]

If the unit cell is defined to be centered at the $1a$ `atomic' position (bounded by black dashed lines in Fig.~\ref{fig:SSH}(a)), then $W^v_0$ would have support on two unit cells. Crucially, if the primitive unit cell is defined to be centered at the 
$1b$ `midbond' position (bounded by green dashed lines in Fig.~\ref{fig:SSH}(a)), then $W^v_0$ can be continuously deformed to lie within said  unit cell. To appreciate this, observe from Fig.~\ref{fig:SSH}(c) that $W^v_0$ currently has support only on the right and left edge of a $1b$-centered unit cell. Since the two orbitals on the right-hand side of \q{RHS} are not individually inversion-symmetric, their centers are not fixed to the $1a$ Wyckoff position. Instead the two orbitals  are mutually related by inversion about the $1b$ position. By a continuous transformation of type (e-ii), one can symmetrically displace the centers of
$\{\varphi'_{R,s\pm p}\}$ 
to  all lie at $1b$ positions; for example,  Fig.~\ref{fig:SSH}(c) illustrates the left-shifting of $\varphi'_{0,s-p}$ and the right-shifting of $\varphi'_{-1,s_+p}$.
It follows that $W^v_0$ now has support only on the $1b$ position with spatial coordinate $-1/2$, demonstrating that the SSH model is  unicellular.

{In spite of the above examples, the notions of multicellular topological insulators and obstructed atomic limits are not necessarily disjoint, and finding an example that simultaneously manifests both notions deserves a separate investigation.}

\section{Multicellularity of RTP models\label{RTP-multicel}}

In this section, we substantiate the claim from the main text that Wannier functions of a non-trivial RTP phase are necessarily multicellular. The proof is by contradiction, namely we first show that unicellular Wannier functions pose constraints on the Berry curvature in the momentum space, whence the multicellularity of Wannier functions in RTP models will be seen as an immediate consequence. Our discussion is split into two parts. First, we prove in Sec.~\ref{sec:flat-Berry} that Wannier functions in RTP models are multicellular, while remaining agnostic about the particular shape of the Wannier function. Then, in Sec.~\ref{sec:RTP-z-extent} we prove a stronger statement, namely that if the RTP is protected by a rotation symmetry around the $z$-axis, 
then the RTP must necessarily extend over multiple layers in the $z$-direction.

\subsection{Flatness of Berry connection in unicellular models}\label{sec:flat-Berry}

Based on the definition given in Eq.~\eqref{eqn:Wanier-strict-uni} we show that
unicellular Wannier functions $\{W_{\bs{R}}^\nu\}_{\bs{R}\in\textrm{BL},\nu\in\{1,\ldots,\mathcal{V}\} }$ encode Bloch bands with flat Berry connection, $\tr[F]=0$. To see this, first note the corresponding Bloch states 
[cf.~Eq.~\eqref{eq:bloch-to-wannier}]
\begin{eqnarray}
\ket{\psi^\nu_{\bs{k}}} 
&=& \sum_{\bs{R}}\mathrm{e}^{i \bs{k}\cdot\bs{R}} \ket{W^\nu_{\bs{R}}} \nonumber \\
&=& \sum_{\bs{R},\alpha} \mathrm{e}^{i \bs{k}\cdot \bs{R}} c_\alpha^\nu   \ket{\varphi_{\bs{R},\alpha}} \nonumber \\
&=& \sum_{\alpha} \mathrm{e}^{-i\bk\bw_\alpha}c_\alpha^\nu \ket{\Phi_{\bk,\alpha}}\label{eq:from-Wannier-to-Bloch}
\end{eqnarray}
where in the second step we use Eq.~\eqref{eqn:Wanier-strict-uni} and in the last step we used the basis of Bloch states given in Eq.~\eqref{eq:bloch_basis} with the positions of the basis Wannier orbitals $\bw_\alpha$. This allows to define tight-binding Bloch vectors
\begin{equation}
    u^\nu(\bk)=\left(\mathrm{e}^{-i\bk\cdot\bw_1}c_1^\nu, \;\dots\; \mathrm{e}^{-i\bk\cdot\bw_{\mathcal{V}+\mathcal{C}}}c_{\mathcal{V}+\mathcal{C}}^\nu\right)^T
    \label{eq:unicellular_bloch_vector}
\end{equation}
that are the eigenvectors of the tight-binding Hamiltonian in momentum space.
We see that the unicellular assumption imposes the derivative to be $\bm{\nabla}_\bk^{\phantom{\nu}} u^\nu_\alpha(\bk)= -i \bw _\alpha^{\phantom{0}}\mathrm{e}^{-i\bk\cdot\bw_\alpha } c^\nu_\alpha$. From this we conclude for the corresponding Berry curvature\footnote{We remark that this conclusion remains true if one adopts the \emph{generalized} unicellularity [cf.~remark (b) in Sec.~\ref{sec:elaboration}] of the valence Wannier functions. In such case $\ket{W_{\bs{R}}^\nu} = \sum_\alpha c_\alpha^\nu \ket{\varphi_{\bs{R}-\bs{r}^\nu_\alpha,\alpha}}$, where $\bs{r}^\nu_\alpha \in\textrm{BL}$ indicates the various choices of the unit cell for the individual Wannier functions, nevertheless each orbital type ($\alpha$) contributes to any Wannier function $\ket{W_{\bs{R}}^\nu}$ only through a single site (the one corresponding to Bravais vector $\bs{R} - \bs{r}_\alpha^\nu$). Then the last line of Eq.~(\ref{eq:from-Wannier-to-Bloch}) is modified to $\ket{\psi^\nu_{\bs{k}}} = \sum_{\alpha} u^\nu_\alpha (\bk) \ket{\Phi_{\bk,\alpha}} = \sum_{\alpha} \mathrm{e}^{i\bk(\br_\alpha^\nu - \bw_\alpha)}c_\alpha^\nu \ket{\Phi_{\bk,\alpha}}$, and the derivative of the tight-binding Bloch vector becomes $\bm{\nabla}_\bk^{\phantom{\nu}} u^\nu_\alpha(\bk) =  i (\br ^\nu_\alpha - \bw_\alpha)\mathrm{e}^{i\bs{k}\cdot(\bs{r}_\alpha^\nu - \bw_\alpha) } c^\nu_\alpha$ and one obtains Berry curvature 
\begin{equation}
\tr[\bs{F}(\bs{k})] = i \sum_{\nu=1}^{\mathcal{V}}\sum_{\alpha=1}^{\mathcal{V}+\mathcal{C}} \abs{c_\alpha^\nu}^2 \left[(\bs{r}_\alpha^\nu-\bw_\alpha) \times (\bs{r}_\alpha^\nu-\bw_\alpha)\right] = 0,
\end{equation}
which vanishes due to the cross product of each position vector $(\bs{r}^\nu_\alpha-\bw_\alpha)$ with itself. Therefore, \emph{generalized} unicellular Wannier functions also lead to flat Berry connection; in other words, RTP insulators are not representable by generalized unicellular Wannier functions. We remark that for the proof presented in Sec.~\ref{sec:RTP-z-extent} we did not find similar extension to generalized unicellularity -- the unicellular discussion assumed therein must be the ``strict'' one as defined by remark (a) in Sec.~\ref{sec:elaboration}.}
\begin{equation}
\tr[\bs{F}(\bs{k})] = i\sum_\nu \left[\bm{\nabla}_\bk u^{\nu}(\bk)\right]^\dagger\times \bm{\nabla}_\bk u^\nu(\bk) = i\sum_{\nu=1}^{\mathcal{V}}\sum_{\alpha=1}^{\mathcal{V}+\mathcal{C}} \abs{c_\alpha^\nu}^2 \left(\bw_\alpha \times \bw_\alpha\right) = 0,
\label{eq:berry_curv_flat}
\end{equation}
which vanishes due to the cross product of each position vector $\bw^\nu_\alpha$ with itself.

It follows from the flatness of the Berry connection [Eq.~(\ref{eq:berry_curv_flat})] that 
unicellular Wannier functions cannot exhibit RTP. To see this, recall that the RTP invariant corresponds to an integral of Berry curvature on certain half-sheet inside the Brillouin zone. As the RTP feature is invariant under continuous deformations that preserve the rotation symmetry and the bulk energy gap, it follows from Eq.~(\ref{eq:berry_curv_flat}) that RTP vanishes for all models that are deformable into a representation with 
unicellular Wannier functions.
Therefore, non-trivial RTP implies multicellular Wannier functions [i.e., ones not compatible with the ansatz in Eq.~(\ref{eqn:Wanier-strict-uni})].

Before concluding this section, we briefly remark that the same conclusion can also be reached by adapting the homotopy argument presented in the main text for the multicellularity of models with non-trivial Hopf invariant. To that end, note that the classifying space of Hamiltonians in the considered symmetry class at a generic momentum $\bs{k}$ is $M = \mathsf{U}(\mathcal{C}+\mathcal{V})/\mathsf{U}(\mathcal{C})\times \mathsf{U}(\mathcal{V})$, where $\mathsf{U}(N)$ is the group of rank-$N$ unitary matrices, and as usual $\mathcal{C}$ ($\mathcal{V}$) denotes the number of conduction (valence) bands. Furthermore, for momenta along high-symmetry lines exhibiting mutually disjoint eigenvalues of the valence vs.~conduction bands, the classifying space (i.e., space of spectrally normalized Hamiltonians) constitutes a single point $(-\mathbf{1}_{\mathcal{V}\times \mathcal{V}})\oplus (+\mathbf{1}_{\mathcal{C}\times \mathcal{C}})\equiv H_0$ in $M$. We now use homotopy theory to study equivalence classes of maps from a sheet bounded by two such high-symmetry lines to $M$, while subject to the constraint that the boundary of the sheet is mapped to $H_0$. It can be shown~\cite{Sun:2018} that these equivalence classes are captured by the pointed homotopy group $\pi_2(M) = \mathbb{Z}$, and that non-trivial RTP corresponds to non-trivial elements of this group.  Since unicellular Wannier functions (i.e., the atomic limits) clearly correspond to the trivial element of this classification, it follows from the homotopy theory that non-trivial elements (i.e., those exhibiting RTP) cannot be continuously deformed into the atomic limit (i.e., they are multicellular).

\subsection{Multi-layered Wannier functions in RTP models}\label{sec:RTP-z-extent}

In this section we prove a \emph{stronger} statement concerning the multicellularity of Wannier functions in rotation-symmetry-protected RTP models: the Wannier functions must extend over several layers along the rotation axis (without loss of generality set along the $z$-direction). To prove this statements, we find it useful to decompose all vectors into an in-plane (perpendicular to the rotation axis, ``$\perp$'') and out-of-plane (parallel with the rotation axis, ``$\parallel$'') components, $\bs{b} = (\bs{b}_\perp,b_\parallel) = \bs{b}_\perp + b_\parallel \hat{\bs{e}}_z$. 
 
We depart from the following ansatz for single-layer Wannier functions, which generalizes the unicellular ansatz in Eq.~(\ref{eqn:Wanier-strict-uni}), 
\begin{equation}
\ket{W^\nu_{\bs{R}}} = \sum_{\alpha,\bs{r}_\perp} c^\nu_{\bs{r}_\perp,\alpha} \ket{\varphi_{\bs{R}-\bs{r}_\perp,\alpha}}\label{eq:Wanier-single-layer}
\end{equation}
where we sum only over ``in-plane'' Bravais vectors $\bs{r}=(\bs{r}_\perp,0)$ with vanishing parallel component. The idea behind the ansatz in Eq.~(\ref{eq:Wanier-single-layer}) is that the support of orbital $\alpha$ to all Wannier functions $\left\{\ket{W^\nu_{\bs{R}}}\right\}_{\nu\in\{1,\ldots,\mathcal{V}\}}$ is limited to a \emph{single layer} labelled by $R_\parallel$ and located at position $\bw_{\alpha,\parallel}$ within unit cell in $z$-direction, 
compatible with the intended meaning of ``single-layer Wannier functions''.

Starting with the Wannier functions of the form in Eq.~(\ref{eq:Wanier-single-layer}), we show that there exists a continuous deformation of the Hamiltonian that preserves the symmetries and the energy gap such that the Berry curvature of the corresponding Bloch states \emph{becomes} oriented along the $z$-direction in the entire Brillouin zone at the end of the deformation. From this we argue that this prevents the realization of RTP models by single-layer Wannier functions.

Proceeding in steps, we first perform the Fourier transformation to obtain the Bloch functions
\begin{eqnarray}
\ket{\psi^\nu_{\bs{k}}} 
&=& \sum_{\bs{R}}\mathrm{e}^{i \bs{k}\cdot\bs{R}} \ket{W^\nu_{\bs{R}}} \nonumber \\
&=&  \sum_{\bs{R}_\perp,R_\parallel,\alpha,\bs{r}_\perp} \mathrm{e}^{i \bs{k}\cdot (\bs{R}_\perp + R_\parallel\hat{\bs{e}}_z)} c^\nu_{\bs{r}_\perp,\alpha} \ket{\varphi_{\bs{R}_\perp+R_\parallel \hat{\bs{e}}_z-\bs{r}_\perp ,\alpha}} \nonumber \\ 
&=&  \sum_{\bs{R}'_\perp,R_\parallel,\alpha,\bs{r}_\perp} \mathrm{e}^{i \bs{k}\cdot (\bs{R}'_\perp +\bs{r}_\perp + R
_\parallel\hat{\bs{e}}_z)} c^\nu_{\bs{r}_\perp,\alpha} \ket{\varphi_{\bs{R}'_\perp+R_\parallel \hat{\bs{e}}_z,\alpha}}  \nonumber\\ 
&=& \sum_{\bs{R}',\alpha,\bs{r}_\perp} \mathrm{e}^{i \bs{k}\cdot \bs{R}'}\mathrm{e}^{i \bs{k}\cdot \bs{r}_\perp } c^\nu_{\bs{r}_\perp,\alpha} \ket{\varphi_{\bs{R}',\alpha}} \nonumber \\  
&=& \sum_{\alpha,\bs{r}_\perp} \mathrm{e}^{i \bs{k}\cdot (\bs{r}_\perp - \bw_\alpha) } c^\nu_{\bs{r}_\perp,\alpha} \ket{\Phi_{\bs{k},\alpha}}
\end{eqnarray}
where in the second line we decomposed $\bs{R}= \bs{R}_\perp + R_\parallel \hat{\bs{e}}_z $, in the third line we substituted $\bs{R}'_\perp = \bs{R}_\perp - \bs{r}_\perp$, and in the fourth line we combined into a single variable $\bs{R}' = \bs{R}'_\perp + R_\parallel\hat{\bs{e}}_z$. In the last line we introduced again the basis for Bloch states from Eq.~(\ref{eq:bloch_basis}).

From the last expression above we can read the tight-binding Bloch vectors, and their derivatives in momentum space
\begin{eqnarray}
u_\alpha^\nu(\bk) &=& \sum_{\bs{r}_\perp} \mathrm{e}^{i \bs{k}\cdot (\bs{r}_\perp -\bw_\alpha^{\phantom{0}}) } c^\nu_{\bs{r}_\perp,\alpha}  \\
\bs{\nabla}^{\phantom{\nu}}_{\bs{k}} u_\alpha^\nu(\bk) &=&  i \sum_{\bs{r}_\perp} \mathrm{e}^{i \bs{k}\cdot (\bs{r}_\perp - \bw_\alpha^{\phantom{0}}) } (\bs{r}_\perp - \bw_\alpha^{\phantom{0}}) c^\nu_{\bs{r}_\perp,\alpha}.
\end{eqnarray}
We can now express the single-band Berry curvature
\begin{eqnarray}
\bs{F}^\nu(\bs{k}) &=& i \left[\bs{\nabla}_\bk u^{\nu}(\bk)\right]^\dagger\times \bm{\nabla}_\bk u^\nu(\bk) \nonumber  \\ 
&=& i \sum_{\alpha,\bs{r}_{\perp 1},\bs{r}_{\perp 2}} \mathrm{e}^{i\bs{k}\cdot\left(\bs{r}_{\perp 2} -\bs{r}_{\perp 1}\right)} \left(\bs{r}_{\perp 1} \times \bs{r}_{\perp 2} - \bw_\alpha \times \bs{r}_{\perp 2} - \bs{r}_{\perp 1} \times \bw_\alpha + \bw_\alpha \times \bw_\alpha\right) \left(c^\nu_{\bs{r}_{\perp1},\alpha}\right)^* c^\nu_{\bs{r}_{\perp 2},\alpha} \nonumber  \\ 
&=& i \sum_{\alpha,\bs{r}_{\perp 1},\bs{r}_{\perp 2}} \mathrm{e}^{i\bs{k}\cdot\left(\bs{r}_{\perp 2} -\bs{r}_{\perp 1}\right)} \left[\bs{r}_{\perp 1} \times \bs{r}_{\perp 2} - \bw_{\alpha,\perp} \times \bs{r}_{\perp 2} - \bs{r}_{\perp 1} \times \bw_{\alpha,\perp} - w_{\alpha,\parallel}(\hat{\bm{e}}_z \times \bs{r}_{\perp 2} + \bs{r}_{\perp 1} \times \hat{\bm{e}}_z) \right] \left(c^\nu_{\bs{r}_{\perp1},\alpha}\right)^* c^\nu_{\bs{r}_{\perp 2},\alpha} 
\label{eq:this-does-not-look-promising}
\end{eqnarray}
where in the last line we separated the orbital position into parallel and orthogonal to the rotation axis components $\bw_\alpha=\bw_{\alpha, \perp} + w_{\alpha,\parallel}\hat{\bm{e}}_z$ and used the fact that a cross product of $\bw_\alpha$ with itself is zero. In the final expression the first three terms give contribution only to the $z$-component of the Berry curvature $F^\nu_z(\bk)$ as they contain cross products of pairs of \emph{in-plane} vectors. At the same time two last terms can be made to vanish by a continuous deformation of the Hamiltonian that preserves the symmetries and the energy gap. In this deformation the centers of the basis orbitals are adiabatically shifted along $z$-axis until $\bw_{\alpha,\parallel}=0$ for all $\alpha$.To preseve the energy gap this must be done while preserving the hoppings between every pair of basis orbitals. After such a deformation we find $\bs{F}^\nu(\bs{k}) \parallel \hat{\bs{e}}_z$ ,and the same is true after we sum over all occupied bands $\nu$. Since the RTP invariant is equal to an integral of Berry curvature on a sheet \emph{parallel} with $\hat{\bs{e}}_z$ and it does not change under continuous deformations of the Hamiltonian described above, the Berry curvature in Eq.~(\ref{eq:this-does-not-look-promising}) cannot generate RTP. It therefore follows that models with a non-trivial RTP invariant cannot be deformed to acquire the Wannier representation in Eq.~(\ref{eq:Wanier-single-layer}), i.e., their Wannier functions are necessarily multi-layered in the $z$-direction.

\section{Non-analyticity of the charge polarization}

In this section we explain the non-analyticity observed in the dependence of the charge polarization $\overline{\mathscr{P}}$ on the tuning parameter $m$ [main text Fig.~2(c,d)], and relate this to the critical points of the presented RTP model [main text Eq.~(1)]. Instead of considering a model periodic in the Brillouin zone, we pretend in this section that the $\bs{k}$-space is infinite in all three directions. This, on the one hand, simplifies the analytic expressions, while on the other hand, results in certain subtleties in properly defining the charge polarization.

We depart from a representative $\bk\cdot\boldsymbol{p}$ expansion near the critical point that changes the returning Thouless pump invariant, $z_1 = (k_x + i k_y)$ and $z_2 = (k_z + i m)$. We further adopt the cylindrical coordinates, $k_x = k \cos \phi$ and $k_y = k\sin\phi$, such that the spinor encoding the Hamiltonian becomes \begin{equation}
z = \left(\begin{array}{c}k \mathrm{e}^{i\phi} \\ k_z + m\end{array}\right),
\end{equation}
and the expanded form of the Hamiltonian is 
\begin{equation}
H = \left(\begin{array}{cc}
k^2 - k_z^2 - m^2   &   \mathrm{e}^{i\phi}k(k_z - im) \\
\mathrm{e}^{-i\phi}k(k_z + im)  & -k^2 + k_z^2 - m^2\label{eqn:k.p-RTP-transition}
\end{array}\right)
\end{equation}
where we dropped the dependence on $(k,\phi,k_z;m)$ for brevity. The valence state can be written in a globally continuous gauge 
\begin{equation}
u^v_0=\frac{i\sigma_y z^*}{\norm{z}} = \frac{1}{\sqrt{k^2 + k_z^2 + m^2}}\left(\begin{array}{c}k_z - i m \\ -\mathrm{e}^{-i\phi}k \end{array}\right)\label{eqn:val}
\end{equation}
whenever $\norm{z}\neq 0$ (which corresponds to the sole gapless point of the model, $k_x=k_y=k_z=m=0$). By taking the derivative of the state in the last equation, we can compute the Berry connection and by integrating it along the $z$-direction we obtain the charge polarization $\mathscr{P}$. 

While the outlined procedure would eventually yield the correct result for the non-analyticity of the polarizations, it exhibits a problem with the physical interpretation: while the Hamiltonian in Eq.~(\ref{eqn:k.p-RTP-transition}) obeys $\lim_{k_z\to\infty}H(k_,\phi,k_;m) = \lim_{k_z\to-\infty}H(k_,\phi,k_;m)$ for any finite $k$, thus allowing us to ``compactify'' the $z$-direction into a closed circle, the gauge for $u^v_0$ in Eq.~(\ref{eqn:val}) does not obey this constraint. Therefore, one cannot readily interpret the constructed integral of the connection as the polarization. 

To overcome the described problem, we augment the expression in Eq.~(\ref{eqn:val}) as
\begin{equation}
u^v=\frac{\mathrm{e}^{i \zeta(k_z)}}{\sqrt{k^2 + k_z^2 + m^2}}\left(\begin{array}{c}k_z - i m \\ -\mathrm{e}^{-i\phi}k  \end{array}\right)\label{eqn:val-gauge}
\end{equation}
where $\zeta(k_z)$ is a smooth function that has limits $\lim_{k_z\to-\infty}=0$ and $\lim_{k_z\to+\infty}=\pi$. Such a choice of $u^v$ is ``smooth at infinity'' and thus can be compactified in the $z$-direction, allowing us to interpret the integral of the connection as the charge polarization. We obtain 
\begin{equation}
A_z = -i \braket{u^v}{\partial_z u^v} = \frac{m}{k^2 + k_z^2 + m^2} + \partial_z \zeta(k_z) \label{eqn:k.p-charge-connection}
\end{equation}
We now explicitly compute the polarization and analyze its non-analyticity. By integrating the connection in Eq.~(\ref{eqn:k.p-charge-connection}) along the (compactified) $z$-direction, we obtain
\begin{equation}
\mathscr{P} = \frac{1}{2\pi}\int_{-\infty}^{+\infty}A_z d k_z = \frac{1}{2}\left(1+\frac{m}{\sqrt{k^2+m^2}}\right).
\end{equation}
Notice that at $k=0$, the polariation jumps from $0$ to $2\pi$ at the critical point as $m$ changes sign from negative to positive, representating the change in the RTP invariant. To compute the total polarization of the Wannier function, the assumed infinite extent of momentum space in $k_{x,y}$ directions requires us to subtract some background from $\mathscr{P}_z$, else we found obtain infinite quantities. Note that such subtraction of a constant does not influence the resulting derivative with respect to $m$. We define the background to be $\mathscr{P}_0 = \frac{1}{2}\left[\lim_{m\to 0^-}\mathscr{P} + \lim_{m\to 0^+}\mathscr{P} \right]$, which for our models is independent of $(k,\phi)$ and equal to $1/2$, and consider the difference $\widetilde{\mathscr{P}}=\mathscr{P}-\mathscr{P}_0$.

The total polarization assuming a 2D system with only $(k_x,k_z)$ coordinates is
\begin{equation}
\mathscr{P}_\textrm{tot.}^\textrm{2D} = \int_{-K}^{+K} dk \,   \widetilde{\mathscr{P}} = \frac{m}{2} \log\left[1+\frac{2K\left(K+\sqrt{m^2 +K^2}\right)}{m^2}\right]
\end{equation}
where the cut-off momentum $K$ 
has to be introduced to keep the integrals finite. The cutoff is most naturally interpreted as the momentum scale characterizing the validity of the effective Hamiltonian with linearized $z(k)$, and is bounded by the size of the Brillouin zone.
For the derivative with respect to the tuning (mass) parameter, we obtain
\begin{equation}
\frac{d}{dm}\mathscr{P}_\textrm{tot.}^\textrm{2D} = \frac{1}{2} \log\left[1+\frac{2K\left(K+\sqrt{m^2 +K^2}\right)}{m^2}\right] - \frac{K}{\sqrt{m^2 + K^2}},
\end{equation}
which diverges at $m\to 0$. In particular, one can approximate the above results for small $m$ as 
\begin{equation}
\frac{d}{dm}\mathscr{P}_\textrm{tot.}^\textrm{2D} \approx \frac{1}{2} \log\left(\frac{4K^2}{m^2}\right),
\end{equation}
from which the logarithmic divergence is apparent. We expect the logarithmic divergence in the change of $\mathscr{P}_\textrm{tot.}^\textrm{2D}$ in two-dimensional models to be indicative of a change of the RTP invariant.

We similarly consider the total polarization of a 3D system,
\begin{equation}
\mathscr{P}_\textrm{tot.}^\textrm{3D} = \int_{0}^{K} 2\pi k\, dk \,   \widetilde{\mathscr{P}} =  m\pi \left(\sqrt{m^2 + K^2} - \abs{m}\right)
\end{equation}
where $K$ is again the cut-off in momentum space, and $\abs{m}$ is the absolute value of $m$. The presence of the absolute value is already indicative of a non-analytic behavior, which is revealed by computing the derivative
\begin{equation}
\frac{d}{dm}\mathscr{P}_\textrm{tot.}^\textrm{3D} = \pi\frac{2m^2+K^2}{\sqrt{m^2+K^2}}-2\pi\abs{m}. 
\end{equation}
We thus conclude that a cusp in the change of $\mathscr{P}_\textrm{tot.}^\textrm{3D}$ in three-dimensional models is indicative of a change of the RTP invariant. Note that in the main text, the finite size of the momentum space allows us to consider a rescaled version of the total polarization, $\overline{\mathscr{P}} = \mathscr{P}_\textrm{tot.}^\textrm{3D}/\textrm{Area}(\textrm{rBZ})$, which ehixibits the same type of non-analyticity. The advantage of the normalization is that if $\mathscr{P}(\bs{k}_\perp)$ is uniquely defined modulo integers, then $\overline{\mathscr{P}}$ is also uniquely defined modulo integers.

\section{Stability of returning Thouless pump under addition of unicellular bands}\label{sec:RTP-stability}

\begin{figure}
    \centering
    \includegraphics{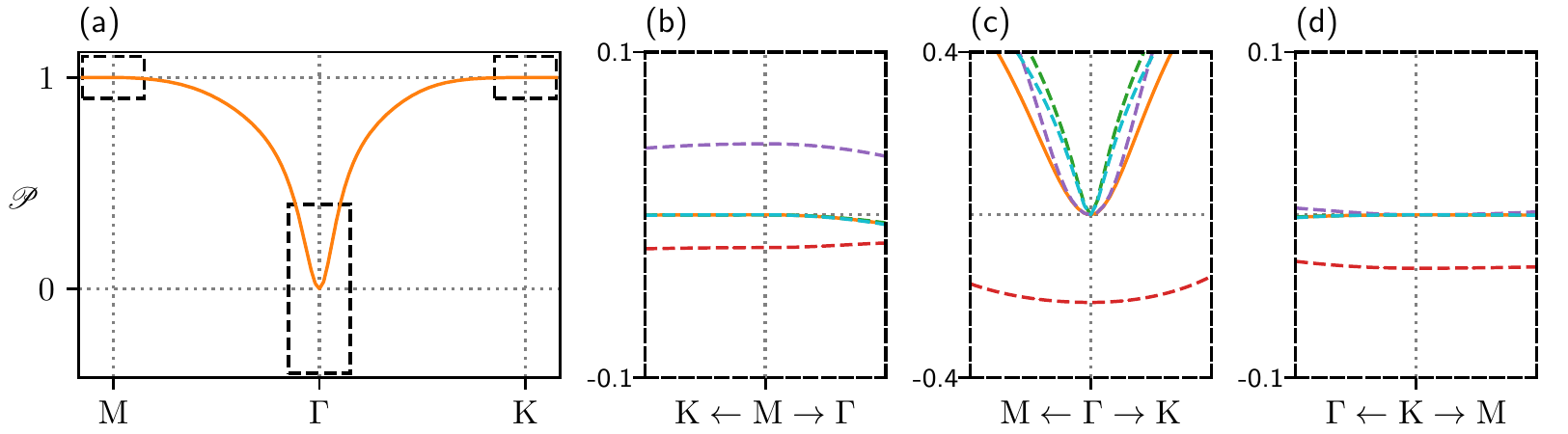}
    \caption{(a) RTP for the two-band model of Eq.~\eqref{eq:hopfc6ham} of the main text at $m=-6$. Three zoom-in rectangles are shown, which correspond to panels (b--d), where RTP of the two-band model is compared against RTP of multi-band models described in Sec.~\ref{sec:RTP-stability}. Solid orange line indicates the original two-band model~\eqref{eq:hopfc6ham} with angular momenta $\ell_v=0$ and $\ell_c=1$. Three-band models with additional conduction band with angular momenta $\ell_c'=1$, $\ell_c'=0$, and $\ell_c'=2$ [captured by Eqs.~(\ref{eq:3-band}--\ref{eqn:3-band-H-lc=2})] are plotted, respectively, by dashed green/red/purple lines. The RTP of a four-band model [Eqs.~(\ref{eq:4-band}--\ref{eqn:4-band-matrix-elems})] with additional valence band with angular momentum $\ell_v'=0$ and additional conduction band with $\ell_c'=1$ is displayed with by dashed blue line.
    }
    \label{fig:RTP-stability}
\end{figure}

We numerically analyze the stability of RTP under addition of a unicellular conduction band whose Wannier representatives transform in one-dimensional representations of rotation. As a starting model that possesses an RTP we use the minimal model \eqref{eq:hopfc6ham} of the main text with a parameter value $m=-6$, which corresponds to $\Delta\mathscr{P}_{\tm\Gamma}=-1$ and $\Delta\mathscr{P}_{\tk\Gamma}=-1$. Its returning Thouless pump (RTP) is shown in Fig.~\ref{fig:RTP-stability}(a) by solid orange line. To better track the changes in polarization, we focus in Fig.~\ref{fig:RTP-stability}(b-d) on the small neighborhoods of the rotation-invariant points $\textrm{M}$, $\Gamma$ and $\textrm{K}$, respectively, which correspond to dashed rectangles in Fig.~\ref{fig:RTP-stability}(a). The non-minimal models with one additional band are described with the following tight-binding Hamiltonian:
\begin{equation}
    H_{3} = \left(\begin{array}{c | c}
    H(\bk)  &  \begin{array}{c}
                h_{vc'}(\bk) \\
                h_{cc'}(\bk) 
               \end{array} \\ \hline 
    \begin{array}{cc}
    h_{vc'}(\bk)^* & h_{cc'}^*(\bk) 
    \end{array} & E(\bk)
    \end{array}\right),
    \label{eq:3-band}
\end{equation}
where $H(\bk)$ the minimal model in Eq.~\eqref{eq:hopfc6ham}, $E(\bk)$ is the energy dispersion of the added band (excluding inter-band hybridization), and $h_{vc'}(\bk)$ [resp.~$h_{cc'}(\bk)$] describes the coupling between the added band and the valence $s$-type (resp.~conduction $p_+$-type) band of the original model. In all tests, an on-site potential is chosen for the added orbital such that $E(\bk)=50$, which is smaller than the bandwidth $\Delta$ of the minimal model {($\Delta=338$, defined as the difference between the maximum energy of the conduction band and the minimum energy of the valence band)}. By fixing the angular momentum $\ell_c'$ of the added band we impose the following constraint on the Hamiltonian
\begin{equation}
    R_{C_6}H_3(\bk)R^{-1}_{C_6} = H_3(C_6\bk), \quad R_{C_6} = \begin{pmatrix}
    1 & 0 & 0 \\
    0 & e^{i2\pi/6} & 0 \\
    0 & 0 & e^{i2\pi\ell_c'/6}
    \end{pmatrix}.
    \label{eq:3band_contraint}
\end{equation}

First, we add a conduction band which transforms in the same representation as the original conduction band,i.e.~$\ell_c'=1$. The symmetry constraint \eqref{eq:3band_contraint} is fulfilled by setting 
\begin{subequations}
\begin{eqnarray}
    h_{vc'}(\bk) &=& 1.5(\cos k_z+2i\sin k_z)\cdot f_{-1}(k_x,k_y), \\
    h_{cc'}(\bk) &=& 1.5i\cdot f_0(k_x,k_y), \\
    f_t(k_x, k_y) &=& \textstyle\sum_{a=1}^{6}\exp(i\pi ta/3)\cdot \exp{i\left[\cos(\pi a/3)k_x+ \sin(\pi a/3)k_y\right]}.\label{eqn:ft-function}
\end{eqnarray}
\end{subequations}
Such a band does not hybridize with the valence subspace along the rotation-invariant lines of the Brillouin zone, thus preserving the quantization of RTP [green dashed line in Fig.~\ref{fig:RTP-stability}(b-d)]. In contrast, when the additional conduction band has $\ell_c'=0$, matching the eigenvalue of the valence band, the quantization of RTP is lost at all high-symmetry points [red dashed line in Fig.~\ref{fig:RTP-stability}(b-d)]. This is achieved by setting
\begin{subequations}
\begin{eqnarray}
    h_{vc'}(\bk) &=& 3(2\cos k_z - 3i\sin k_z)\cdot f_0(k_x,k_y), \\
    h_{cc'} (\bk) &=& 0
\end{eqnarray}
\end{subequations}
with $f_0(k_x,k_y)$ defined by Eq.~(\ref{eqn:ft-function}) above.

When the representation of the additional conduction band is set to $\ell_c'=2$ by choosing
\begin{subequations}\label{eqn:3-band-H-lc=2}
\begin{eqnarray}
    h_{vc'}(\bk) &=& 3(\cos k_z + 4i\sin k_z)\cdot f_{-2}(k_x,k_y) , \\
    h_{cc'}(\bk) &=& 3(\cos k_z + 3i\sin k_z) \cdot f_{-1}(k_x,k_y),
\end{eqnarray}
\end{subequations}
its angular momentum coincides with that of the valence subspace at $C_2$-invariant points. {This reflects the fact that if we view the model as $C_2$-symmetric (forgetting its $C_6$ symmetry), the mutually-disjoint condition is not satisfied, hence the polarization difference along $\Gamma\textrm{M}$ is no longer integer-quantized.
On the other hand, if the model is viewed as $C_3$-symmetric, the mutually-disjoint condition is satisfied, hence} the RTP along $\Gamma\textrm{K}$ remains quantized [purple dashed line in Fig.~\ref{fig:RTP-stability}(b-d)]. 

Additionally, we consider a four-band model where bands added to both valence and conduction subspaces keep the mutually-disjoint condition, having $l_v'=0$ and $l_c'=1$. It is given by a tight-binding Hamiltonian
\begin{equation}
    H_4(\bk) = \left(\begin{array}{c|c|c}
         -E(\bk) & 
         \begin{array}{cc}
         h_{v'v}(\bk) & h_{v'c}(\bk)
         \end{array}  & 0 \\ \hline
         \begin{array}{c}
         h^*_{v'v}(\bk) \\
         h^*_{v'c}(\bk)
         \end{array} & H(\bk) &
         \begin{array}{c}
         h_{vc'}(\bk) \\
         h_{cc'}(\bk)
         \end{array} \\ \hline
         0 & \begin{array}{cc}
         h^*_{vc'}(\bk) & h^*_{cc'}(\bk)
         \end{array}  & E(\bk)         
    \end{array}\right),
    \label{eq:4-band}
\end{equation}
with the same choice of $E(\bk) = 50$, and with further matrix elements
\begin{subequations}\label{eqn:4-band-matrix-elems}
\begin{eqnarray}
    h_{v'v}(\bk) &=& \exp(ik_z)f_0(k_x,k_y), \\
    h_{v'c}(\bk) &=& -2(\cos k_z - i\sin k_z)\cdot f_{-1}(k_x,k_y), \\
    h_{vc'}(\bk) &=& (\cos k_z + 2i\sin k_z) \cdot f_{-1}(k_x,k_y), \\
    h_{cc'}(\bk) &=& \exp(ik_z)\cdot f_0(k_x,k_y),
\end{eqnarray}
\end{subequations}
that satisfy the symmetry constraint
\begin{equation}
    R_{C_6}H_4(\bk)R^{-1}_{C_6} = H_4(C_6\bk), \quad R_{C_6} = \begin{pmatrix}
    1 & 0 & 0 & 0 \\
    0 & 1 & 0 & 0 \\
    0 & 0 & e^{i2\pi/6} & 0 \\
    0 & 0 & 0 & e^{i2\pi/6}
    \end{pmatrix}.
    \label{eq:4band_contraint}
\end{equation}
We observe that the RTP remains quantized [blue dashed line in Fig.~\ref{fig:RTP-stability}(b-d)] under the addition of unicallular valence and conduction bands that respect the mutually-disjoint condition. The presented study of RTP stability illustrates the notion of symmetry-protected delicate topology. 

\section{RTP-Hopf mod-six correspondence}

We present a complete proof of the RTP-Hopf mod-six correspondence [Eq.~(\ref{eq:RTP-hopf-C6}) of the main text], which has only been sketched in the main text. We remind the reader that the correspondence holds for any $C_6$-symmetric, Pauli-matrix (i.e., two-band) Hamiltonian with trivial first Chern class and with the property that representative Wannier functions of the valence and conduction bands are centered on the $C_6$-symmetric Wyckoff position, and transform (under $C_6$) with angular momenta $\ell_v=0$ and $\ell_c=1$, respectively. (A follow-up work~\cite{Nelson:2021} will present generalized correspondence for any $C_n, \ell_v, \ell_c$ and Wyckoff positions.) Our proof utilizes an equivalent formula for the Hopf invariant derived by Whitehead~\cite{Whitehead:1947}, which we briefly review. 

\subsection{Review of Whitehead formulation of the Hopf invariant}
\label{sec:whitehead}

We consider a Hamiltonian $H(\bk)$ with energy gap $\varepsilon^c(\bk)-\varepsilon^v(\bk)>0$ at each three-momentum $\bk$, i.e., 
\begin{equation}
H(\bk) = \ket{u^c(\bk)}\varepsilon^c(\bk)\bra{u^c(\bk)} + \ket{u^v(\bk)}\varepsilon^v(\bk)\bra{u^v(\bk)},
\end{equation}
and we perform a continuous deformation of the energies to $\varepsilon^c(\bk) = +1$ and $\varepsilon^c(\bk) = -1$. 
Each such a Pauli-matrix (i.e., two-band) Hamiltonian can be interpreted as a map from the Brillouin zone (BZ) to a Bloch sphere of \emph{conduction} pseudospin-half vectors {(or, equivalently, into the classifying space of $2$-band Hamiltonians, $\textsf{U}(2)/\textsf{U}(1)\times\textsf{U}(1)\cong S^2$)}. Specifically for Hamiltonians expressed via a (normalized) two-component spinor $z(\bk)$ [cf.~first line of Eq.~(\ref{eq:hopfc6ham}) in the main text], one finds that
\begin{equation}
\boldsymbol{h}:\bk \mapsto {z}^{\dagger}(\bk)\bsigma z(\bk)
\end{equation}
produces a normalized three component vector $\boldsymbol{h}(\bk) \in S^2$. The Hamiltonian corresponds to $H(\bk) = \boldsymbol{\bk}\cdot \boldsymbol{\sigma}$, and the conduction state is $\ket{u^c(\bk)} = z(\bk)$ which indeed corresponds to point $\boldsymbol{\bk}$ on the Bloch sphere. Such a dual description is not surprising due to the relation between the conduction state and the (spectrally normalized) Hamiltonian, $H(\bk)=2\ket{u^c(\bk)}\bra{u^c(\bk)}-\mathbf{1}$. We remark, however, that the decomposition of the Hamiltonian into the vector of Pauli matrices encoded by $\boldsymbol{h}(\bk)$ is true generally for any two-band Hamiltonian, even if not expressed via spinor $z(\bk)$.

We further assume that the Hamiltonian map has trivial first Chern class, thus excluding the Hopf-Chern insulators introduced in \ocite{kennedy_hopfchern}. In such a case, the preimage of any point $x_0\in S^2$ on the Bloch sphere is an orientable (but not necessarily path-connected) 1-manifold in BZ
\begin{equation}
\boldsymbol{h}^{-1}(x_0) = \mcup_i\gamma_i,
\end{equation} 
with orientation defined to be anti-parallel to the tangential component of the Berry curvature of the \emph{valence} band (i.e., parallel if instead considering the conduction band).\footnote{Being anti-parallel rather than parallel is a matter of convention.} We call the 1-manifold $\mcup_i \gamma_i$ the \emph{preimage path} of $x_0\in S^2$. 
A possible choice of the components $\gamma_i$ of a preimage path is illustrated in Fig.~\ref{fig:Whitehead}(a) by green lines:  $\gamma_1$ is a closed loop while $\gamma_2$ consists of two non-contractible loops winding around the BZ torus. That the two non-contractible loops have opposite orientation is not an accident, and is a consequence of a pairing rule derived in~\s{sec:threerules}. 

Whitehead showed~\cite{Whitehead:1947} that the Hopf invariant coincides with the Chern number $\mathscr{C}$ on the oriented open Gaussian surface $\Sigma$ (so-called \emph{Seifert surface}) whose boundary is the preimage path, $\partial\Sigma=\mcup_i\gamma_i$, i.e.
\begin{equation}
    \chi=\mathscr{C}_\Sigma:=\frac{1}{2\pi}\int_{\Sigma}\bm{\mathcal{F}}\cdot d\bm{\Sigma},
    \label{eq:whitehead}
\end{equation}
with $\bm{\mathcal{F}}$ the Berry curvature of the \emph{valence} band. (From now on, whenever we mention Berry curvature without further specification, we have in mind the one corresponding to the valence band.) Note that the Chern number is well-defined \emph{and quantized}, because the wave function of the filled band is constant on the preimage $\mcup_i \gamma_i$, hence the surface $\Sigma$ can be treated as closed. 

\subsection{Proof of the modulo-six RTP-Hopf correspondence}

For the class of $C_6$-invariant Hamiltonians considered in the main text, we choose, without loss of generality, 
a basis such that the $\ell=0$ (resp.\ $\ell=1$) state {corresponds to the expectation value} $\langle\sigma_z\rangle{=}1$ ($\langle\sigma_z\rangle{=}{-}1$). The symmetry constraint on the Hamiltonian is then
\begin{equation}
R_{C_6}H(\bk)R_{C_6}^{-1}=H(C_6\bk),\;\;\;\; R_{C_6}=\begin{pmatrix}1 & 0 \\ 0 & \exp(i2\pi /6). \end{pmatrix}\la{c6constraint}
\end{equation}
By assumption, the valence-band energy eigenvector at all rotation-invariant lines of the BZ belongs to the one-dimensional $\ell=0$ subspace, hence all these lines belong to the preimage of the south pole on the Bloch sphere, $(0,0,-1)\in S^2$.\footnote{Note that notations $\boldsymbol{h}^{-1}[(0,0,-1)]$ and $H^{-1}(-\sigma_z)$ for the preimage can be used interchangeably. Below we mostly use the latter.} We denote the $C_6,C_3,C_2$-invariant lines by $\gamma_\Gamma$, $\{\gamma_\tk, \gamma_{\tk'}\}$ and $\{\gamma_\tm, \gamma_{\tm'}, \gamma_{\tm''}\}$, respectively, as illustrated by green lines in Fig.~\ref{fig:Whitehead}(b). Note that lines $\gamma_{\tk,\tk'}$ and $\gamma_{\tm,\tm',\tm''}$ are mutually related by $C_6$ rotation. 

\subsubsection{Assigning orientations to preimage loops}\la{sec:threerules}

We assign orientation to all $\{\gamma_{\bkp'}\}$ in accordance with the following rules:  \\

\noi{i}  \textit{Pairing rule}: any intersection of {a 2D BZ subtorus} with the preimage (of any single point on the Bloch sphere) must come in pairs with opposite orientation. \\

\noi{ii} If  $\gamma_{\bkp'}$ and $\gamma_{\bkp''}$ are related by rotation, they must have the same orientation (owing to the Berry curvature transforming under spatial transformations as a pseudovector). \\

\noi{iii} The orientations of $\gamma_{\Gamma,\tk,\tk'}$ point upwards (as derived from the difference between conduction and valence angular momenta, $\ell_c-\ell_v = 1$, in a paragraph below).\\
\bigskip 

\noindent \textit{Derivation of pairing rule (i).---} Our assumption of a trivial first Chern class means that the first Chern number vanishes on any 2D cut of the BZ, and in particular it vanishes for all 2D subtori $T^2$ of the BZ. Parametrizing $T^2$ by $(k_x,k_y)$ and defining $k_z$ by the right-hand rule, the Chern number is given by the following integral of the Berry curvature: $\mathscr{C}=(2\pi)^{-1}\int_{T^2} \bm{\mathcal{F}}_z dk_xdk_y$. Recall that the first Chern number $\mathscr{C}$ of a pseudospinor Hamiltonian defined on  $T^2$  tells how many times this manifold wraps around the Bloch sphere under the Hamiltonian map $H: T^2 \ri S^2$. If $\mathscr{C}=0$, then any point  $x_0$ on the Bloch sphere must be visited an even number of times as one sweeps through $T^2$, i.e., the preimage of $x_0$ consists of an even number of points $\{\bk_1,\ldots ,\bk_{2N}\}$ in $T^2$. Let us define $d_{x_0}\in S^2$ as an infinitesimal {outward (i.e., positively) oriented} disk centered at ${x_0}$, and {$\delta\Omega_{x_0}>0$} as the solid angle subtended by $d_{x_0}$. The preimage of $d_{x_0}$ comprises  $2N$ disks  $\{D_1,\ldots ,D_{2N}\}$ encircling $\{\bk_1,\ldots ,\bk_{2N}\}$, respectively, {with orientations inherited from $T^2$}. We now assign each $D_j$ an index, $\textrm{ind} [D_j]=\pm 1$, by checking whether $H(\partial D_j)$ (with the orientation of the boundary determined by right-hand rule) is parallel or antiparallel to the oriented boundary $(\partial d_{x_0})\subset S^2$ [i.e., whether it winds (counter-)clockwise around $x_0$]. Applying the solid-angle interpretation of the Berry phase for pseudospinor Hamiltonians~\cite{berry_quantalphase}, the Berry curvature integrated over {the oriented disk $D_j$} equals {$\textrm{ind}[D_j] \delta \Omega_{x_0}/2$}. (This equality also manifests that {$-\textrm{ind}[D_j]=-\sgn[\mathcal{F}_z(\bk_j)]$}, which is the definition of the preimage orientation adopted in Sec.~\ref{sec:whitehead}.) Because $d_{x_0}$ is covered by the map $(H: T^2 \ri S^2)$ a number of times equal to the Chern number (assumed zero), the net Berry curvature integrated over $\{D_1,\ldots ,D_{2N}\}$ must vanish, implying that $N$ of the disks $\{D_1,\ldots ,D_{2N}\}$ have {index} opposite to the remaining $N$ disks. This completes the proof.
\bigskip

\begin{figure}
    \centering
    \includegraphics{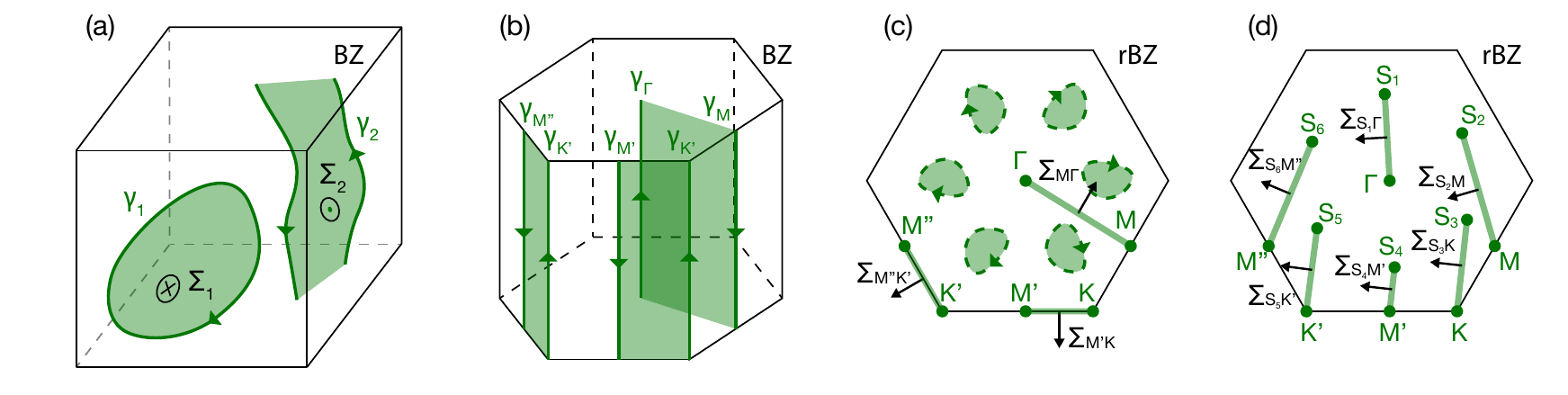}
    \caption{(a) Oriented preimage $H^{-1}(x_0)=\gamma_1\cup\gamma_2$ inside the Brillouin zone (BZ) of a point $x_0$ on a Bloch sphere, assuming a generic two-band (Pauli-matrix) Hamiltonian. The sheet $\Sigma_1\cup\Sigma_2$ is an oriented surface stretched over the preimage, with orientation defined using the right-hand rule. (b) In a {$C_6$-invariant model with $\ell_c=1$ and $\ell_v=0$}, the rotation-invariant lines $\gamma_\Gamma$, $\gamma_\tk$, $\gamma_{\tk'}$, $\gamma_\tm$, $\gamma_{\tm'}$ and $\gamma_{\tm''}$ are preimages of the south pole on the Bloch sphere. Their orientation is defined anti-parallel to the {tangent component of} Berry curvature of the valence band. (c) Projection into the reduced Brillouin zone (rBZ); this includes the projection of surfaces $\Sigma_{\tm\Gamma}$, $\Sigma_{\tm'\tk}$ and $\Sigma_{\tm''\tk'}$ stretched over the six rotation invariant lines, with orientations indicated by the black arrows.
    Additional possible preimages that appear in {six copies related by $C_6$} are illustrated. (d) Additional preimages in the form of six non-contractible lines (shown straight in the illustration, but they need not have this property in general) are denoted by $S_i$ in the projection to rBZ. Surfaces that are stretched over the full preimage are denoted as $\Sigma_{\ts_1\Gamma}$, $\Sigma_{\ts_3\tk}$, $\Sigma_{\ts_5\tk'}$, $\Sigma_{\ts_2\tm}$, $\Sigma_{\ts_4\tm'}$ and $\Sigma_{\ts_6\tm''}$.}
    \label{fig:Whitehead}
\end{figure}

\noindent \emph{Derivation of rule (iii).---} This follows from the assumed angular momenta ($\ell_v=0$ and $\ell_c=1$) and from a $\bk\cdot \boldsymbol{p}$ analysis at the $C_3$-invariant wavevectors.  (Note, however, that the little groups of $\{\tm,\tm',\tm''\}$ are not sufficiently constraining to determine  the orientations of $\gamma_{\tm,\tm',\tm''}$.) Since the preimage orientation depends only on the Berry curvature (a property of the wave function), it may as well be determined by the spectrally-flattened Hamiltonian.
By assumption, such a `flat-band' Hamiltonian at rotation-invariant lines is $H=\mathbf{1}-2(1,0)^\top(1,0)=- \sigma_z$, {with $\mathbf{1}$ the identity matrix}. Moving slightly away from $\gamma_\Gamma$ (or $\gamma_{\tk,\tk'}$), the leading-order correction to this Hamiltonian is determined from \q{c6constraint} to be
\begin{equation}
H(\bk)=-h_+\sigma_+-h_-\sigma_--h_z\sigma_z, \as (h_+,h_-,h_z)=(ak_-,a^*k_+,1),\as k_\pm=k_x\pm ik_y, \as     \sigma_\pm=\sigma_x\pm i\sigma_y,\as a\in \mathbb{C}.\la{hj}
\end{equation} 
In general, the filled-band Berry curvature of a two-by-two, flat-band Hamiltonian, $H(\bk)= -\sum_{i=1}^3 q_i(\bk)\sigma_i$ with $||\boldsymbol{q} ||=1$, is expressible  as a skyrmion density~\cite{qi2005},
\begin{equation}
\mathcal{F}_z =-\frac{1}{2}\epsilon_{ijk} q_i \partial_x q_j \partial_y q_k.
\end{equation}
In the particular case that $q_3$ is independent of $k_x$ and $k_y$, the general expression reduces to
\begin{equation}
\mathcal{F}_z =-\frac{1}{2}q_3 (\partial_x q_1 \partial_y q_2-\partial_x q_2 \partial_y q_1)= 2h_z(\partial_+h_+\partial_-h_- - \partial_+h_-\partial_-h_+)+ O(\bkp^2), \as \partial_{\pm}=\f1{2}\big(\partial_{k_x}\mp i\partial_{k_y}\big).\label{eqn:Berry-algebra}
\end{equation}
In the last step, we substituted $h_z=q_3+O(\bkp^2)$ and $2h_{\pm}=q_1\mp iq_2 +O(\bkp^2)$, with $O(\bkp^2)$ terms resulting from having to normalize $||\boldsymbol{q} ||=1$. Substituting the expressions for $\{h_j\}_{j\in\{+,-,z\}}$ from~\q{hj} into Eq.~(\ref{eqn:Berry-algebra}), we obtain $\mathcal{F}_z=-2|a|^2 + O(\bkp^2)$. The orientation of the preimage (at $C_3$-invariant $\bk$) is defined to be anti-parallel to $(0,0,\mathcal{F}_z)$, hence the three \emph{upward}-facing arrows at $\gamma_{\Gamma},\gamma_{\textrm{K}},\gamma_{\textrm{K}'}$ displayed in  Fig.~\ref{fig:Whitehead}(b). [Note that if we chose a different basis in which $\ell=0$ (resp.\ $\ell=1$) state has expectation $\langle\sigma_z\rangle{=}{-}1$ ($\langle\sigma_z\rangle{=}1$), while fixing the valence subspace to have zero angular momentum, then an analogous symmetry analysis gives $(h_+,h_-,h_z)=(a^*k_+,ak_-,-1)$, which gives the same value for $\mathcal{F}_z$, and hence also the same orientation for the preimage at $C_3$-invariant $\bk$.]

\subsubsection{Assuming the south-pole preimage comprises only rotation-invariant lines}\label{sec:simplified-proof}

Postponing the more general situation to a {subsubsection below}, let us first assume that the south-pole preimage comprises only the rotation-invariant lines, as is indeed true for the minimal model in the main text. These lines intersect the $k_z=0$ subtorus at six points, allowing us to apply rule (i)  from Sec.~\ref{sec:threerules}. Combining with rule (iii), we deduce that the all three $\gamma_\tm$ lines are \emph{downward} oriented [cf.\ Fig~\ref{fig:Whitehead}(b)], which is also consistent with rule (ii). A possible choice for oriented surfaces bounded by the six $\gamma$ lines are the three light-green sheets in Fig.~\ref{fig:Whitehead}(b), which we denote by $\Sigma_{\tm\Gamma}$, $\Sigma_{\tm'\tk}$ and $\Sigma_{\tm''\tk'}$; the orientation of these surfaces are determined by the right-hand rule, as shown by black arrows in Fig.~\ref{fig:Whitehead}(c). The Chern number contributed by the oriented surface $\Sigma_{\bk'_\perp\bk''_\perp}$ is the difference in polarization between upward- and downward-oriented edges, $\mathscr{C}_{\bkp'\bkp''} = \mathscr{P}_{\bk''_\perp}-\mathscr{P}_{\bk'_\perp}$, assuming that the polarization is continuously defined over the reduced Brillouin zone (rBZ). Finally, applying the Whitehead formula in Eq.~\eqref{eq:whitehead}, and the equality of $\mathscr{P}$ at symmetry-related $\bkp'$, we derive the exact equality: $\chi=2\Delta\mathscr{P}_{\tm\tk}+\Delta\mathscr{P}_{\tm\Gamma}=3\Delta\mathscr{P}_{\tm\Gamma}-2\Delta\mathscr{P}_{\tk\Gamma}$. This corresponds to Eq.~(\ref{eq:RTP-hopf-C6}) of the main text with the mod-six `$\equiv_6$' relation replaced by an exact equality.

\subsubsection{Assuming the south-pole preimage comprises more than the rotation-invariant lines}\label{sec:complete-proof}

If the south-pole preimage comprises more than the rotation-invariant lines, then the RTP-Hopf relation is generalized to Eq.~(\ref{eq:RTP-hopf-C6}) with the mod-six condition re-instated.
\begin{equation}
    \chi\equiv_6 2\Delta\mathscr{P}_{\tm\tk}+\Delta\mathscr{P}_{\tm\Gamma}.
    \label{eq:rtp-hopf}
\end{equation}
There are five classes of possibilities for additional preimages, of which the last three involve a nontrivial linking of two or more preimage loops $\gamma_i$:\\

\begin{itemize}
\item[(i)] The additional preimages can form contractible loops as in Fig.~\ref{fig:Whitehead}(c), which appear in six copies related by $C_6$ symmetry. \\

\item[(ii)] Additional preimages can wind around the BZ in the form of six rotation-related lines $\gamma_{\ts_i}$ as illustrated in Fig.~\ref{fig:Whitehead}(d). To satisfy the pairing rule,  the orientations of $C_2$-invariant lines $\gamma_\tm$ must be upward-oriented, while all $\gamma_\ts$ lines are downward-oriented. The oriented surfaces which contribute to Hopf invariant are  $\Sigma_{\ts_1\Gamma}$, $\Sigma_{\ts_3\tk}$, $\Sigma_{\ts_5\tk'}$, $\Sigma_{\ts_2\tm}$, $\Sigma_{\ts_4\tm'}$ and $\Sigma_{\ts_6\tm''}$, as illustrated in Fig.~\ref{fig:Whitehead}(d). \\

\item[(iii)] A preimage loop may link with (i.e., encircle) the non-contractible $\gamma_{\Gamma}$-loop.\\

\item[(iv)] A  preimage loop may link with the $\gamma_{\tk}$-loop, alongside a $C_2$-related loop that links with the $\gamma_{\tk'}$-loop.\\

\item[(v)] A preimage loop may link with the $\gamma_{\tm}$-loop, with a $C_3$-related (resp.\ $C^{-1}_3$-related) loop linking with the $\gamma_{\tm'}$-loop (resp.\ $\gamma_{\tm''}$-loop).\\

\end{itemize}

\noindent Let us prove \q{eq:rtp-hopf} for cases (i--v) in turn. \bigskip

For case (i), note that the loops arise in six $C_6$-related copies. For symmetry reasons, the oriented surface stretched over each of these six loops carry the same Chern number $\mathscr{C}'$, such that together they contribute $6 \mathscr{C}'$ to the Hopf invariant. This reduces the exact equality to the mod-six relation in Eq.~(\ref{eq:rtp-hopf}).
\bigskip

For case (ii), the six non-contractible loops at generic wavevectors are generally curvilinear [they are depicted in Fig.~\ref{fig:Whitehead}(d) for better clarity of the illustration]. The parallel transport of Bloch wave functions along each non-contractible loop defines a Zak phase $\phi_Z(\ts_j)$ (as the line integral of the Berry-Zak connection); the geometric theory of polarization gives $\phi_Z(\ts_j)/2\pi \equiv_1 \mathscr{P}(\ts_j)$ \emph{only for straight} non-contractible loops. {However,} observe that the eigenvector of $H(\bk)$ is constant along $\gamma_{\ts_j}$, with said constant vector being the zero-angular-momentum state. It thus follows that the Zak phase reduces~\cite{TBO_JHAA} to $\phi_Z/2\pi \equiv_1\boldsymbol{G}\cdot \boldsymbol{w}$, with  $\boldsymbol{G}$ the reciprocal vector connecting the intersection of $\gamma_{\ts_j}$ with the BZ boundary, and $\boldsymbol{w}$ the central position of a representative, zero-angular-momentum Wannier function. We have already established in the main text that $\mathscr{P}(\bkp')\equiv_1\boldsymbol{G}\cdot \boldsymbol{w}$, thus $\mathscr{P}(\bkp')-\phi_Z(\ts_j)/2\pi \in \mathbb{Z}$. The six-fold symmetry guarantees that $\phi_Z(\ts_j)$ is independent of $j$, assuming the Bloch wave function is analytic and periodic over the Brillouin zone -- a condition readily satisfied because of the triviality of the first Chern class. 

\begin{figure}
    \centering
    \includegraphics[width=0.99\textwidth]{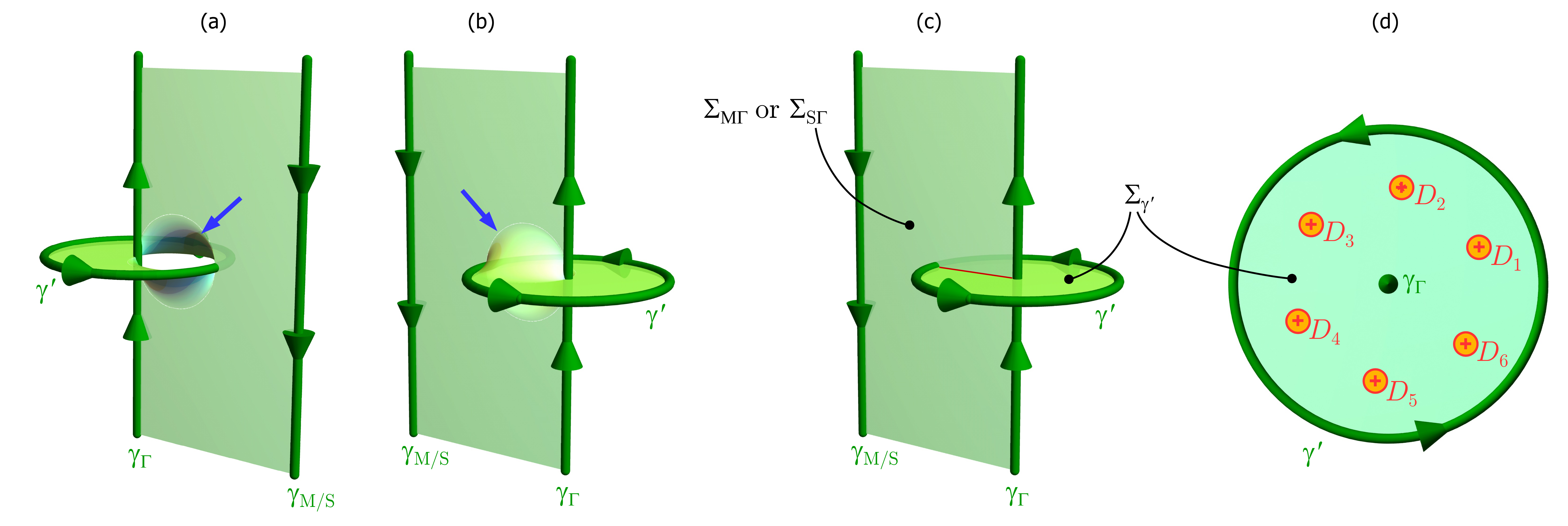}
    \caption{{Illustrations for  case (iii) of additional preimages, as listed in Sec.~\ref{sec:complete-proof}. All oriented lines are components of $\mcup_i\gamma_i = H^{-1}(-\sigma_z)$.  (a,b) Two views (back vs.~front) of a smooth open Gaussian surface (pale green) bounded by non-contractible paths $\gamma_\Gamma$ and $\gamma_\textrm{M/S}$, and by a contractible path $\gamma'$ that is linked with $\gamma_\Gamma$ (oriented green lines). The inscribed sheet is flat everywhere, except for a small inflection region indicated by blue arrow. (c) By continuously shrinking the inflection region, one can effectively deform the inscribed surface into a pair of flat sheets (labelled $\Sigma_{\textrm{M}\Gamma}$/$\Sigma_{\textrm{S}\Gamma}$ and $\Sigma_{\gamma'}$) that cross along the red line. (d) The Chern number on $\Sigma_{\gamma'}$ (viewed from the projected $z$-direction) is determined by counting the indices $\textrm{ind}[D_j]$ (set to ``$+$'' in the figure) of preimages of a small neighborhood of the \emph{south} pole ${+}\sigma_z{\in}S^2$. Due to $C_6$-symmetry, these appear in multiples of six, $\{D_j\}_{j=1}^6$ (orange disks), with the same index. It follows that the Chern number $\mathscr{C}_{\gamma'}$ is an integer multiple of six.}
    }
    \label{fig:helicoid}
\end{figure}

Applying the Whitehead formula, as well as the equality of polarization (or the Zak phase) for $C_6$-related paths, we obtain 
\begin{equation}
    \chi=\mathscr{P}(\Gamma) + 2\mathscr{P}(\tk) + 3\mathscr{P}(\tm)-6\phi_Z(\ts)/2\pi.\la{chiint}
\end{equation} 
The integer quantization of $\mathscr{P}(\tm)- \phi_Z(\ts)/2\pi$ allows to substitute  $\left[3\mathscr{P}(\tm)-6\phi_Z(\ts)/2\pi\equiv_6-3\mathscr{P}(\tm) \right]$ in \q{chiint}, leading to the desired relation in \q{eq:rtp-hopf}.
\bigskip

For case (iii) with a preimage loop $\gamma'$ encircling $\gamma_{\Gamma}$, a smooth open Gaussian surface bounded by $\gamma'$ and $\gamma_{\Gamma}$ [Fig.~\ref{fig:helicoid}(a,b)] may be continuously deformed, and then split into two intersecting surfaces [Fig.~\ref{fig:helicoid}(c)]: one, denoted $\Sigma_{\gamma'}$, being $C_6$-symmetric and bounded by $\gamma'$ \textit{alone}, and the other surface being $C_6$-asymmetric and bounded on one side by $\gamma_{\Gamma}$ [more precisely, it is either $\Sigma_{\textrm{M}\Gamma}$ in Fig.~\ref{fig:Whitehead}(c) or $\Sigma_{\textrm{S}_1\Gamma}$ in Fig.~\ref{fig:Whitehead}(d)]. The contribution to $\chi$ by the second surface has already been analyzed in cases (i) and (ii) above, resp.~in Sec.~\ref{sec:simplified-proof}. It thus remains to prove that $\Sigma_{\gamma'}$ can only contribute an integer multiple of six to $\chi$.

{To prove that Chern number $\mathscr{C}_{\gamma'}$ on the $C_6$-symmetric surface $\Sigma_{\gamma'}$ is quantized to integer multiples of six, we utilize the concepts developed while deriving the pairing rule in Sec.~\ref{sec:threerules}. We remind the reader that $\gamma'$ belongs to the preimage of the south pole ($-\sigma_z$) on the Bloch sphere; it is convenient to also consider the preimage of the \emph{north} pole, $H^{-1}(+\sigma_z)$. Owing to the matrix representation of $C_6$ being simultaneously diagonal with $\sigma_z$, it follows that both $H^{-1}( \sigma_z)$ and $H^{-1}(- \sigma_z)$  are $C_6$-symmetric. We emphasize that the $C_6$ symmetry extends to the orientations of the discussed preimages, because of the pseudovector transformation of the Berry curvature. (We remark however that $C_6$ symmetry is generically \emph{not} a property of the preimages of other points $x_0 {\in} S^2$.) It follows from the discussion in Sec.~\ref{sec:threerules} that $\mathscr{C}_{\gamma'}=\sum_j \textrm{ind}[D_j]$, where $D_j\subset \Sigma_{\gamma'}$ are preimages of a small neighborhood $d_{-\sigma_z}\subset S^2$ of the south pole. As both $\Sigma_{\gamma'}$ and $H^{-1}(-\sigma_z)$ are $C_6$-symmetric, it follows that the preimages $D_j$ (which cannot lie at $\gamma_\Gamma$ because $H(\gamma_\Gamma){=}{-}\sigma_z$) come in multiples of six, with all members of the sextuplet having the same index [Fig.~\ref{fig:helicoid}(d)]. As a consequence, the contribution of the preimage $\gamma'$ to $\chi$ is $\mathscr{C}_\gamma'\equiv_6 0$, thus preserving the validity of Eq.~(\ref{eq:rtp-hopf}).}
\bigskip

{The cases (iv) and (v) are analyzed with analogous arguments as those presented above for (iii). In the case (iv) of a pair of loops $\gamma_{1,2}'$ encircling $\textrm{K}$ resp.~$\textrm{K}'$, one ends up considering a pair of surfaces $\Sigma_{\gamma_{1,2}'}$, each stretched along one of the two loops. It follows from the symmetry that the Chern numbers on the two surfaces are $\mathscr{C}_{\gamma_1'}=\mathscr{C}_{\gamma_2'} \equiv_3 0$. Therefore, their net contribution to $\chi$ is a multiple of six, thus preserving the validity of Eq.~(\ref{eq:rtp-hopf}). Similarly, the three symmetry-related loops $\gamma_{1,2,3}'$ that arise in case (v) lead us to consider three symmetry-related open Gaussian surfaces $\Sigma_{\gamma_{1,2,3}'}$, each carrying $\mathscr{C}_{\gamma_i'}\equiv_2 0$. It again follows that the net contribution of the additional preimages to $\chi$ is a multiple of six.}

\section{Strong obstruction principle for the Hopf insulator}

The strong obstruction principle for the Hopf insulator states that there is no exponentially-localized Wannier representation for the Hilbert space of states defined on a half-infinite slab. We have claimed that this Hilbert space includes all states, independent of their filling and spatial extension. The meaning of this Hilbert space will be precisely established  here, to complement the heuristic description given in the main text. Once the meaning is established, we will be able to prove the strong obstruction principle with greater rigor.

\subsection{Proof of obstruction principle}

{By assumption of the triviality of the first Chern class, any surface-localized band (if it exists) can always be removed from the Fermi level by a deformation of the surface Hamiltonian. This implies the existence of an energy gap} separating a filled subspace ({defining the projector} $P$) and unfilled subspace (with orthogonal projector $Q$). We then consider Bloch-Wannier eigenstates of $P\hat{z}P$ and $Q\hat{z}Q$, with eigenvalues of $\hat{z}$ taking only positive values.  We will see that adopting the Bloch-Wannier representation is not just a convenient choice of basis, it also allows to define the Hilbert space on a semi-infinite geometry. \\

We label the ``filled'' Bloch-Wannier eigenbands of $P\hat{z}P$ by an index $b{=}1,2,\ldots,b_{\textrm{max}}-1,b_{\textrm{max}},b_{\textrm{max}}+1,\ldots$, such that band $b$ lies closer (to the surface termination) than  band $b'$, if $b<b'$. We impose that $b_{\textrm{max}}$ is sufficiently large, such that the Bloch-Wannier band with the same index is bulk-like, i.e., it is indistinguishable (up to exponentially small corrections) from a bulk Bloch-Wannier band defined with periodic boundary conditions. In particular, this means that  band $b_{\textrm{max}}$ is related to $b_{\textrm{max}}{\pm} 1$ by a discrete translation mapping $z\ri z\pm 1$. We define the filled Hilbert space $\calh_{P}[b_{\textrm{max}}]$ as  the set of Bloch-Wannier bands labelled by $b{=}1,2,\ldots,b_{\textrm{max}}$. By similar consideration of the ``unfilled'' eigenbands of $Q\hat{z}Q$, we define the unfilled Hilbert space $\calh_{Q}[b_{\textrm{max}}]$ with the same truncation $b_{\textrm{max}}$. The full Hilbert space of states on a half-infinite geometry is given by  $\calh_{1/2}=\calh_{P}\oplus \calh_{Q}$, with $b_{\textrm{max}}$ taken sequentially to infinity. This procedure of defining an infinite-dimensional  Hilbert space by sequential embeddings in increasingly larger Hilbert spaces is not unlike the direct-limit procedure employed in $K$-theory~\cite{kitaev_periodictable}.\\

We then compute the Chern number $\mathscr{C}_P(b)$ of each band as an integral of the Berry curvature over the $\textrm{rBZ}$, and define the sum $\mathscr{C}_P[B]{=}\sum_{b=1}^B \mathscr{C}_P(b)$. Viewed as a sequence in $B$, $\mathscr{C}_P[B]$ has a unique accumulation point (defined as $\mathscr{C}_P$) for large enough $B$ (satisfying $B<b_{\textrm{max}}$), because all bulk Bloch-Wannier bands have trivial Chern number owing to the bulk translational symmetry;  $\mathscr{C}_P$ has the physical meaning of the Chern number of filled Bloch-Wannier bands localized to a finite vicinity of the surface. We analogously define $\mathscr{C}_Q$ as the Chern number of unfilled, surface-localized Bloch-Wannier bands. The net Chern number of all surface-localized bands, \textit{independent} of filling, is then $\mathscr{C}_f{=}\mathscr{C}_P+\mathscr{C}_Q$. This \textit{faceted Chern number} $(\mathscr{C}_f)$ equals the bulk invariant $\chi$, according  to the bulk-boundary correspondence proven in \ocite{AA_teleportation}. Crucially $\mathscr{C}_f$ is the net Chern number of the entire Hilbert space $\calh_{1/2}[b_{\textrm{max}}]$ for $b_{\textrm{max}}$ that is sufficiently large (in the sense described above). The relation $\chi=\mathscr{C}_f\neq 0$ thus implies there exists no  exponentially-localized Wannier representation of $\calh_{1/2}[b_{\textrm{max}}]$, for any large $b_{\textrm{max}}$; in particular, this means that no such representation exists as we take $b_{\textrm{max}}\ri\infty$ in the above-described direct-limit procedure.

\subsection{Delicacy of obstruction principle}

{When  a unicellular, bulk conduction band is added to the Hopf insulator, the Wannier obstruction described in the previous subsection no longer holds for all values of the truncation parameter $b_{max}$. Instead, the existence of an obstruction depends on the parity of $b_{max}$; for one parity, we find that the obstruction is removable.}

The addition of a unicellular bulk conduction band implies there are two bulk-like, unfilled Bloch-Wannier bands in any interval $[z,z+1]$, for $z$ that is sufficiently far from the surface termination.  The net Chern number of both bulk-like bands vanishes, in accordance with the triviality of the first Chern class in the bulk. However, the two bulk-like bands can have cancelling Chern numbers; by a continuous deformation of $Q$, it is always possible that one bulk-like band has Chern number $-\mathscr{C}_f$ and the other has Chern number $+\mathscr{C}_f$. In defining the unfilled Hilbert space $\calh_Q[b_{\textrm{max}}]$, we see that advancing $b_{\textrm{max}}$ by one changes the net Chern number of  $\calh_Q$ by ${\pm} \mathscr{C}_f$.  Thus there exists $b_{\textrm{max}}$ of one parity such that the bulk-like Bloch-Wannier bands have a net Chern number $-\mathscr{C}_f$ that cancels the Chern number of the topologically-nontrivial surface bands -- this implies that $\calh_{1/2}[b_{\textrm{max}}]$ has trivial Chern number and possesses an exponentially-localized Wannier representation. In contrast, for  $b_{\textrm{max}}$ of the opposite parity $\calh_{1/2}[b_{\textrm{max}}]$ remains topologically nontrivial.

\section{Symmetry-indicator analysis of surface bands of the \texorpdfstring{$P3$}{P3}-symmetric RTP insulator}

By analysis of the symmetry representations in $\bk$-space, we identify which of the possible, rank-two surface bands (of the $P3$-symmetric RTP insulator) are compatible with the symmetry representations of a band representation. We further assume that a $P3$-symmetric band -- with trivial first Chern class and the symmetry representations of a band representation -- is identifiable with said band representation. (Exceptions to this rule are known to exist for certain space groups~\cite{crystalsplit_AAJHWCLL}, but not for the $Pn$ groups, with $n=2,3,4,6$.)

We assume that angular momentum $\ell=1$ corresponds to point-group representation $^2\!{E}$ (and also to little-group representations $\Gamma_3$, $\textrm{K}_3$, $\textrm{K}'_2$), while $\ell=2$ corresponds to $^1\!{E}$ (and to $\Gamma_2$,  $\textrm{K}_2$, $\textrm{K}'_3$), and we use the \texttt{BANDREP} tool on the Bilbao crystallographic server~\cite{elcoro_EBRinBilbao} to find the decompositions for the surface bands of the three-band semi-infinite model.

\subsection{Detailed analysis of the bottom surface band}
In this subsection we provide more technical details for the discussion in the main text. The surface discussed there corresponds to a semi-infinite slab defined for $z>0$ meaning that it is a bottom surface of a slab. The symmetry indicators of the corresponding surface bands are presented in Fig.~\ref{fig:bbc}(c) in the main text.

First, $\textrm{SB}_1^{\phantom{\alpha}}\oplus\textrm{SB}_2^\alpha$ is obstructed, meaning it is not decomposable into elementary band representations. That this obstruction is fragile can be proven by adding $s$ orbitals on $1b$ and $1c$ Wyckoff positions as
\begin{equation}
\textrm{SB}_1^{\phantom{\alpha}}\oplus\textrm{SB}_2^\alpha\oplus[A_1\uparrow G]_{1b} \oplus[A_1\uparrow G]_{1c} = [(2A_1\oplus{^1}\!E\oplus{^2}\!E)\uparrow G]_{1a}.\label{eqn:SB-SB1a}
\end{equation}
In contrast, combinations $\textrm{SB}_1^{\phantom{\beta}}\oplus\textrm{SB}_2^\beta$ and $\textrm{SB}_1^{\phantom{\gamma}}\oplus\textrm{SB}_2^\gamma$ are decomposable into elementary band representations, but neither combination satisfies the uniaxial condition with both Wannier centers on the $1a$ position. In the former case, one of the two Wannier centers lies on the $1c$, while in the latter case it is $1b$:
\begin{subequations}\label{subeqn:SB2-bc}
\begin{eqnarray}
\textrm{SB}_1^{\phantom{\beta}}\oplus\textrm{SB}_2^\beta &=& [A_1 \uparrow G]_{1a}\oplus [{^2}\!E \uparrow G]_{1c}\\
\textrm{SB}_1^{\phantom{\gamma}}\oplus\textrm{SB}_2^\gamma &=& [A_1 \uparrow G]_{1a}\oplus [{^2}\!E \uparrow G]_{1b}.
\end{eqnarray}
\end{subequations}
Similar to the decomposition of $\textrm{SB}_1^{\phantom{\alpha}}\oplus\textrm{SB}_2^\alpha$ in Eq.~(\ref{eqn:SB-SB1a}), the direct sums in Eqs.~(\ref{subeqn:SB2-bc}) can be composed with elementary band representations corresponding to Wyckoff positions $1b$ or $1c$ (thus violating that uniaxial condition), such the the resulting bands are Wannier-representable with orbitals residing solely on the $1a$ Wyckoff position, namely:
\begin{subequations}\label{subeqn:SB2-bc-1a}
\begin{eqnarray}
\textrm{SB}_1^{\phantom{\beta}}\oplus\textrm{SB}_2^\beta \oplus [(A_1\oplus {^1}\!E)\uparrow G]_{1c} &=& [(2A_1 \oplus {^1}\!E\oplus {^2}\!E) \uparrow G]_{1a}\\
\textrm{SB}_1^{\phantom{\gamma}}\oplus\textrm{SB}_2^\gamma \oplus [(A_1\oplus {^1}\!E)\uparrow G]_{1b} &=& [(2A_1 \oplus {^1}\!E\oplus {^2}\!E) \uparrow G]_{1a}.
\end{eqnarray}
\end{subequations}

\subsection{Symmetry-indicator analysis of the top surface band}

To complete the discussion presented in the main text we also study decompositions for surface bands of a slab defined for $z<0$, which thus has a top surface. In this case the nontrivial surface band $\textrm{TSB}_1$ has representations coinciding with the valence bulk band at $\Gamma$, $\ell_\Gamma=0$, and with the conduction bulk band at $\tk$ and $\tkpr$, $\ell_\tk=\ell_\tkpr=1$, as shown in the first row of Tab.~\ref{tab:sym-ind}. This band has a Chern number $\mathscr{C}_f=-1$ and thus is not band representable. As for the bottom surface no band detached from the conduction subspace can nullify the surface Chern number as such band has $\mathscr{C}\equiv_3 0$. Symmetry indicators of all possible bands $\textrm{TSB}_2^{\alpha,\beta,\gamma}$ detached from the valence subspace and having Chern number $\mathscr{C}=1$ are presented in the second to fourth rows of Tab.~\ref{tab:sym-ind}. We see that combination of $\mathrm{TSB}_1$ band with any of these bands is band representable with all Wannier centers being different from the bulk Wannier center $1a$:
\begin{subequations}\label{subeqn:SB2-abc}
\begin{eqnarray}
\textrm{TSB}_1^{\phantom{a}}\oplus\textrm{TSB}_2^\alpha &=& [A_1 \uparrow G]_{1b}\oplus [A_1 \uparrow G]_{1c}\\
\textrm{TSB}_1^{\phantom{b}}\oplus\textrm{TSB}_2^\beta &=& [A_1 \uparrow G]_{1c}\oplus [{^1}\!E \uparrow G]_{1c}\\
\textrm{TSB}_1^{\phantom{c}}\oplus\textrm{TSB}_2^\gamma &=& [A_1 \uparrow G]_{1b}\oplus [{^1}\!E \uparrow G]_{1b}.
\end{eqnarray}
\end{subequations}
The band representations obtained from a Wannier function centered at $1a$ Wyckoff position have the same symmetry indicator at all $C_3$-invariant points. Thus, the surface bands of the multicellular topological insulator that have symmetry indicators of conduction subspace at some but not all $C_3$-invariant points can never form only $1a$ centered Wannier functions.

\begin{table}[h!]
    \centering
    \begin{tabular}{c|c|c|c}
         & $\ell_\Gamma$ & $\ell_\tk$ & $\ell_\tkpr$ \\
         \hline
         $\textrm{TSB}_1$ & 0 & 1 & 1 \\
         \hline
         $\textrm{TSB}_2^\alpha$ & 0 & 2 & 2 \\
         $\textrm{TSB}_2^\beta$ & 2 & 0 & 2 \\
         $\textrm{TSB}_2^\gamma$ & 2 & 2 & 0
    \end{tabular}
    \caption{Angular momenta of top surface bands (TSB) of a semi-infinite slab}
    \label{tab:sym-ind}
\end{table}

\section{\label{sec:finite_slab}Finite slab models}

\subsection{\label{sec:fanite_slab_matrix}Hamiltonian for a slab geometry}
Here we present a finite slab Hamiltonian which was used to obtain the spectrum in Fig.~\ref{fig:bbc}(a,b). The system is periodic in $x$ and $y$ spatial directions, and open in the $z$ direction with $N$ layers. For sufficiently large $N$, the surface states of a finite slab (localized to one of the two surface facets) approximates the surface states of a half-infinite slab; the half-infinite geometry plays an important role in the bulk-boundary correspondence  discussed in the main text. 

The Hamiltonian is represented by an $N\times N$ block matrix that retains its dependence on momentum inside the reduced BZ $\bk_\perp=(k_x, k_y)$:
\begin{equation}
    H^r_{\textrm{slab}}(\bk_\perp)= 
    \begin{pmatrix}
        \varepsilon^r & J_1^r & J_2^r & 0 & \dots \\
        J_1^{r,\dagger} & \varepsilon^r & J_1^r & J_2^r  \\
        J_2^{r,\dagger} & J_1^{r,\dagger} & \varepsilon^r & J_1^r  \\
        0 & J_2^{r,\dagger} & J_1^{r,\dagger} & \varepsilon^r \\
        \vdots & & & & \ddots
    \end{pmatrix},
    \label{eq:slab_ham}
\end{equation}
where $\varepsilon^r = \varepsilon^r(\bk_\perp)$ and $J_i^r=J_i^r(\bk_\perp)$ are $r\times r$ blocks of a finite model corresponding to a rank-$r$ bulk Hamiltonian. The block $\varepsilon^r$ describes intra-layer potential while $J_1^r$ ($J_2^r$) describes nearest (next-nearest) neighbor layers coupling. For a minimal two-band model \eqref{eq:hopfc6ham} of the main text they are given by the following matrices:
\begin{subequations}
\begin{eqnarray}
    & \varepsilon^2(\bk_\perp)=
    \begin{pmatrix}
        \left|f_{-1}(\bk_\perp)\right|^2 - \left[f_0(\bk_\perp)+m\right]^2 - 17/2 & -i\left[f_0(\bk_\perp)+m\right]\cdot f_{-1}(\bk_\perp)\\
        i\left[f_0(\bk_\perp)+m\right]\cdot f_{-1}^*(\bk_\perp) & -\left|f_{-1}(\bk_\perp)\right|^2 + \left[f_0(\bk_\perp)+m\right]^2 + 17/2
    \end{pmatrix}, \\
    & J_1^2(\bk_\perp)=
    \begin{pmatrix}
        -4\left[f_0(\bk_\perp)+m\right] & -5i/2\cdot f_{-1}(\bk_\perp) \\
        3i/2\cdot f_{-1}^*(\bk_\perp) & 4\left[f_0(\bk_\perp)+m)\right]
    \end{pmatrix}, \\
    & J_2^2(\bk_\perp)=
    \begin{pmatrix}
        -15/4 & 0 \\
        0 & 15/4
    \end{pmatrix},
\end{eqnarray}
\label{eq:2band_blocks}
\end{subequations}
with $f_t(\bk_\perp)$ given in Eq.~\eqref{eqn:ft-function}. For a rank-three bulk model with additional valence band with angular momentum $\ell_v'=2$ the blocks are given by:
\begin{subequations}
\begin{eqnarray}
    & \varepsilon^3(\bk_\perp)=\left(\begin{array}{c|c}
         -50 & \begin{array}{cc} 0.4 f_2(\bk_\perp) & 0.56f_1(\bk_\perp) \end{array} \\
         \hline
         \begin{array}{c}
         0.4 f_2^*(\bk_\perp) \\
         0.56f_1^*(\bk_\perp)
         \end{array}
         & \varepsilon^2(\bk_\perp)
    \end{array}
    \right), \\
    & J_1^3(\bk_\perp)= \left(\begin{array}{c|c}
         0 & \begin{array}{cc} 0.8 f_2(\bk_\perp) & 0.8f_1(\bk_\perp) \end{array} \\
         \hline
         \begin{array}{c}
         -0.24 f_2^*(\bk_\perp) \\
         0.32f_1^*(\bk_\perp)
         \end{array}
         & J_1^2(\bk_\perp)
    \end{array}
    \right), \\
    & J_2^3(\bk_\perp)= \left(\begin{array}{c|c}
         0 & \begin{array}{cc} 0 & 0 \end{array} \\
         \hline
         \begin{array}{c} 0 \\
         0 \end{array}
         & J_2^2(\bk_\perp)
    \end{array}
    \right).
\end{eqnarray}
\label{eq:3band_blocks}
\end{subequations}
The spectrum in Fig.~\ref{fig:bbc}(a) is calculated for the given two-band system, with $N=50$ layers and parameter value $m=-6$. The spectrum in Fig.~\ref{fig:bbc}(b) is given by a combination of the described three-band system with the same number of layers $N$ and parameter $m$ and an additional layer of 2-bands attached to the lower surface of the slab as described in the following subsection.

\subsection{\label{sec:add_layer} Getting Wannierizable surface states}

In this section we present a model that realizes a 3-band insulator with returning Thouless pump and Wannierizable surface states shown in Fig.~\ref{fig:wannier}(e, f) and Fig.~\ref{fig:bbc}(b). The original 3-band model given in Eq.~\eqref{eq:3band_blocks} possesses a surface state with angular momenta $0$, $1$ and $0$ at momenta $\tk$, $\Gamma$ and $\tkpr$ correspondingly. To be able to Wannierize the surface we need to detach this band from the rest of the spectrum as described in Sec.~\ref{sec:detach} and to peel off another band from the bulk that will nullify the total Chern number of the surface. A band $\textrm{SB}_2^\beta$ from Fig.~\ref{fig:bbc}(c) having angular momenta $2$, $0$ and $0$ works for this purpose. 

In our simulation instead of peeling off such band from the bulk we attach to the surface two additional band representations with angular momenta $0$ and $2$ and perform a band inversion at $\tk$ producing a band with angular momenta $2$, $0$ and $0$ as required and a complementary band with angular momenta $0$, $2$ and $2$ which we shift to large negative energies away from the bulk gap. Such 2-band Hamiltonian takes the following form:
\begin{subequations}
\begin{eqnarray}
    & H_\textrm{layer}(\bk_\perp) = z(\bk_\perp)^\dagger\bsigma z(\bk_\perp)\bsigma + 1.4 f(\bk_\perp) \sigma_z, \\
    & z(\bk_\perp) = (z_1(\bk_\perp), z_2(\bk_\perp))^T, \\
    & z_1(\bk_\perp) = \sum_{a=0}^{2}e^{i4\pi a/3}\cos(
            \cos(2\pi a/3)k_x + \sin(2\pi a/3)k_y), \\
    & z_2(\bk_\perp) = \sum_{a=0}^{2} \exp(i\left[\cos(2\pi a/3)k_x + \sin(2\pi a/3)k_y\right]), \\
    & f(\bk_\perp) = g(\bk_\perp)\left(g(\bk_\perp) - 3\sqrt{3} / 2\right), \\
    & g(\bk_\perp) = \sum_{a=0}^{2} \sin(
            \cos(2\pi a/3)k_x + \sin(2\pi a/3)k_y)
\end{eqnarray}
\label{eq:ham_layer}
\end{subequations}
To simplify surface modifications we additionally perform flattening of these bands as we describe in Sec.~\ref{sec:flattening} and get the intra-layer Hamiltonian $H_\textrm{layer, flattened}(\bk_\perp)$ which we rescale and shift by constant energy to fit the required band inside the bulk gap. Finally, we hybridize the added bands with the outer-most layer of the original Hamiltoian $H_\textrm{slab}^3$ such that the rotational symmetry is preserved. The final Hamiltonian consists of the following blocks
\begin{equation}
    H_\textrm{slab+layer}(\bk_\perp)=
    \left(\begin{array}{c|c}
         H_\textrm{layer, flattened}(\bk_\perp) & \begin{array}{ccc} h_\textrm{hybridization}(\bk_\perp) & 0 & \cdots \end{array} \\
         \hline
         \begin{array}{c}
         h_\textrm{hybridization}^\dagger(\bk_\perp) \\
         0 \\
         \vdots
         \end{array}
         & H_\textrm{slab}^3(\bk_\perp)
    \end{array}
    \right)
    \label{eq:ham_slab+layer}
\end{equation}

\subsection{\label{sec:wannier} Wannierization of the surface states}

The set of eigenvalues of the constructed Hamiltonian $H_{\textrm{slab+layer}}(\bk_\perp)$ possesses two non-degenerate surface-localized states $u_1(\bk_\perp)$ and $u_2(\bk_\perp)$ detached from the rest of the spectrum and having opposite Chern numbers. This allows us to perform a gauge smoothening procedure to obtain two Wannier representable states. For this define the unit cell spanned by the vectors. $\bm{b}_1$, $\bm{b}_2$: $\bk_\perp=k_1\bm{b}_1 + k_2\bm{b}_2$ with $\bR_i\bm{b}_j=\delta_{ij}$, lattice vectors $\bR_1=(1/2, \sqrt{3}/2)$, $\bR_2=(1/2, -\sqrt{3}/2)$ and $k_i\in[0, 2\pi]$.  For each value of $k_1$ define an operator $\hat{\mathcal{W}}_{q_2}(k_1)=\prod_{k_2:\, q_2\leftarrow 0}\ket{u(k_1, k_2)}\bra{u(k_1, k_2)}$ where $u = (u_1, u_2)$. To get a smooth gauge of $u$ we perform the following steps:
\\
\noi{i} At $k_1=0$ take initial vectors $\ket{\omega_j(0,0)}$, $j=1,2$ such that 
\begin{equation}
    \hat{\mathcal{W}}_{2\pi}(0)\ket{\omega_j(0,0)} = \exp(i\theta_j(0))\ket{\omega_j(0,0)}.
    \label{eq:smooth00}
\end{equation}
\noi{ii} For all consequent momenta $k_1$ with a step size $\Delta$ get
\begin{equation}
    \hat{\mathcal{W}}_{2\pi}(k_1)\ket{\tilde{\omega}_j(k_1,0)} = \exp(i\theta_j(k_1))\ket{\tilde{\omega}_j(0,0)}
    \label{eq:smoothk10}
\end{equation} 
and choose smooth vectors for $j=1,2$ as 
\begin{equation}
    \ket{\omega_j(k_1, 0)}=\ket{\tilde{\omega}_j(k_1, 0)}\bra{\tilde{\omega}_j(k_1,0)}\ket{\omega_j(k_1 - \Delta, 0)}.
    \label{eq:smoothk10_proj}
\end{equation}
\noi{iii} To unwind an accumulated phase modify each vector:
\begin{equation}
    \ket{v_j(k_1,0)}=e^{-ik_1/2\pi\; \lambda_j}\ket{\omega_j(k_1, 0)},
    \label{eq:smoothk10_periodic}
\end{equation}
with $\bra{\omega_j(0, 0)}\ket{\omega_j(2\pi, 0)}=e^{i\lambda_j}$.
\\
\noi{iv} To get smooth vectors at all $k_2$ points calculate:
\begin{equation}
    \ket{v_j(k_1, k_2)}=e^{-ik_2/2\pi\; \theta_j(k_1)}\hat{\mathcal{W}}_{k_2}(k_1)\ket{v_j(k_1, 0)}.
    \label{eq:smoothk1k2}
\end{equation}

Obtained vectors $\ket{v_1}$ and $\ket{v_2}$ are smooth and periodic in the BZ but they lost the symmetry of the system. To restore it we apply a symmetrization algorithm described in Ref.~\cite{crystalsplit_AAJHWCLL}. By performing the Fourier transform of these symmetric smooth periodic Bloch functions we get exponentially decaying symmetric Wannier vectors shown in Fig.~\ref{fig:wannier}(e,f) of the main text.

\subsection{\label{sec:flattening}Band flattening}
In this section we describe an algorithm to obtain approximately flat bands. We start with a 2-dimensional 2-band tight-binding Hamiltonian with a gapped spectrum. After finding the spectral decomposition $H(\bk_\perp)=U(\bk_\perp)E(\bk_\perp)U^\dagger(\bk_\perp)$ we consider a spectrally flattened Hamiltonian $H_{flat}(\bk_\perp)=U(\bk_\perp)(-\sigma_z) U^\dagger(\bk_\perp)$ which is topologically equivalent to $H$ and has flat spectrum with energies $\pm 1$. Perfectly flat bands are possible by a cost of infinitely large hoppings making the model non-physical. In a more realistic model we keep a finite number of real-space hoppings which is large enough to have sufficiently flat energy bands. This can be done by performing a Fourier transform of the momentum-space Hamiltonian, truncating all real-space hoppings exceeding some fixed distance (15 unit cells in our case) and finally performing a Fourier transform back to momentum space to get a 2-band Hamiltonian $H_{layer,flattened}(\bk_\perp)$ with required symmetry indicators and almost flat bands.

\subsection{\label{sec:detach}Algorithm to detach a surface state}
Here we outline the algorithm which was used to detach surface bands from the rest of the spectrum in the slab models described in Sec.~\ref{sec:fanite_slab_matrix}. Importantly, we modify Hamiltonian only on the surface thus keeping the bulk unaffected. 

First, we describe the detachment procedure in a two-band model [cf.~Eq.~\eqref{eq:2band_blocks}]. Since we are interested in only one (lower) surface, we completely remove the upper surface state from the gap. To do so we add a potential expressed by a diagonal matrix $V=\textrm{diag}(18, 3)$ to the corresponding intra-layer block and get $H^2_{\textrm{slab};NN}(\bk_\perp)=\varepsilon^2(\bk_\perp){+}V$. To detach the lower surface band from all other bands we reduce the lower-most intra-layer potential by multiplying it with $0.3$ factor: $H^2_{\textrm{slab};11}(\bk_\perp)=0.3\varepsilon^2(\bk_\perp)$. This brings the lower surface state closer to zero energy and detaches it from the bulk bands.

We perform analogous steps with slightly different parameters for a three-band model with added two-band layer [cf.~Eq.~(\ref{eq:3band_blocks}, \ref{eq:ham_slab+layer})], which allows us to detach two surface bands with opposite Chern numbers from the rest of the spectrum. Similar to the two-band case, the upper surface state is removed by taking the upper intra-layer potential $H^3_{\textrm{slab},NN}(\bk_\perp)=\varepsilon^3(\bk_\perp) {+} \textrm{diag}(3,20,3)$. On the lower surface we detach the energy band by reducing the lower-most intra-layer potential of the Hopf slab $H^3_{\textrm{slab};11}(\bk_\perp)=0.3\varepsilon^3(\bk_\perp)$. The second energy band inside the gap originates from the added surface layer as described in Sec.~\ref{sec:add_layer}.

While not essential for the detaching procedure, when producing Fig.~\ref{fig:bbc}(a,b) we performed an additional step of pushing all hybrid bands (which are partially surface-like and partially bulk-like) out of the bulk energy gap. This `pushing' is done by the method of projectors: at each point $(\bk_\perp)\in\textrm{rBZ}$ for the $n^\textrm{th}$ eigenvector $\ket{u_n}$ (ordered according to increasing energy) we define a projector $P_n=\ket{u_n}\bra{u_n}$. Additionally, we define a projector to the top (bottom) layer $P_{t/b}^r$ which has all elements zero except the last (first) $r\times r$ diagonal block, which is equal to the identity matrix. Modification of the slab Hamiltonian $H^r_{\textrm{slab}}\mapsto H^r_{\textrm{slab}}+\alpha^r_n P_{t/b}^rP_nP_{t/b}^r$ with a properly chosen real coefficient $\alpha^r_n$ allows us to project the surface-localized part of the $n^\textrm{th}$ eigenstate to the bulk spectrum. This effectively removes the corresponding energies from the gap. Fig.~\ref{fig:bbc}(a,b) is obtained after performing a series of projections with non-zero coefficients presented in Tab.~\ref{tab:proj_coef} first for a two-band and then for a three-band model.

{\renewcommand{\arraystretch}{1.2}
\begin{table}
    \centering

    \begin{tabular}{l|cc}
          & $P^r_t$ & $P^r_b$ \\
          \hline\hline
         $\alpha^2_{N}$ & $0$ & $20$ \\
         $\alpha^2_{N-2}$ & $-9$ & $-3.6$ \\
         \hline
         $\alpha^3_{N}$ & $0$ & $20$ \\
         $\alpha^3_{N-3}$ & $-15$ & $-5$
    \end{tabular}

    \caption{Coefficients in front of the projectors $P^r_{t/b}P_nP^r_{t/b}$ added to the Hamiltonian to project surface states to the bulk}
    \label{tab:proj_coef}
\end{table}}

\section{Hexagonal magnetic space groups that support Returning Thouless pump}\label{sec:MSGs}


In this section we provide a list of hexagonal magnetic space groups (MSGs) which allow for a non-trivial delicate-topological RTP. Adopting a Euclidean coordinate system $(x,y,z)$ as in the main text, where $z$ lies parallel to the six-fold axis, the necessary criteria are
\begin{itemize}
\item Absence of $\mathcal{PT}: (x,y,z,t)\mapsto (-x,-y,-z,-t)$ (space-time inversion) symmetry and $M_z:(x,y,z,t)\mapsto (x,y,-z,t)$ (horizontal mirror) symmetry, as these quantize polarization at all momenta $\bs{k}_\perp\in\textrm{rBZ}$. This implies $\Delta\mathscr{P}_{\bs{k}',\bs{k}''}\stackrel{!}{=}0$ for all pairs of momenta in the reduced Brillouin zone (rBZ), i.e., the absence of RTP. 
\item Absence of $C_{2x}:(x,y,z,t)\mapsto(x,-y,-z,t)$ and $C_{2y}\mathcal{T}:(x,y,z,t)\mapsto (-x,y,-z,-t)$ symmetry which enforce quantization of polarization along the $\Gamma$--K--M line in the reduced BZ, leading to the absence of RTP.
\item Absence of spatial-inversion $\mathcal{P}:(x,y,z,t)\mapsto(-x,-y,-z,t)$ symmetry {which is incompatible with a delicate RTP phase. The reason is that if certain amount of electric charge is pumped over half a BZ-period, $\mathcal{P}$ symmetry guarantees the same amount of charge is pumped also over the complementary half; as a consequence resulting in a (stably topological) non-vanishing first Chern class.}
\end{itemize}

\begin{table}[t!]
    \centering
    \begin{tabular}{c|ccc|c} \hline \hline
	MSG number (label -- type)  &   \multicolumn{3}{c|}{example pair of EBRs}    & pairs of momenta supporting RTP \\ \hline 
    168.109 ($P6$ -- I)     &   \EBR{A}{1}{1a}    &\adjoin&      \EBR{^1E_2}{1}{1a}         &  \allpoints \\
    168.110 ($P61'$ -- II)  &   \EBR{^1E_1{}^2E_1}{2}{1a}   &\adjoin& \EBR{B}{1}{1a}        &  \allpoints \textsuperscript{(2)} \\
    168.111 ($P6'$ -- III)  &   \EBR{A_1}{1}{1a}   &\adjoin& \EBR{^1E{}^2E}{2}{1a}          &  \GKKprime \textsuperscript{(5)}\\
    168.112 ($P_c6$ -- IV)  &   \EBR{A}{2}{2a}   &\adjoin& \EBR{^1E_2}{2}{2a}               &  \allpoints \textsuperscript{(2)}\\    
    171.121 ($P6_2$ -- I)   &   \EBR{A}{3}{3a}    &\adjoin&      \EBR{B}{3}{3a}             &  $\Gamma$--M \\ \hline
    171.122 ($P6_21'$ -- II)&   \EBR{A}{3}{3a}    &\adjoin&      \EBR{B}{3}{3a}             &  $\Gamma$--M \textsuperscript{(1)}\\
    171.124 ($P_c6_2$ -- IV)&   \EBR{A}{6}{6a}    &\adjoin&      \EBR{B}{6}{6a}             &  $\Gamma$--M \textsuperscript{(1)}\\
    172.125 ($P6_4$ -- I)   &   \EBR{A}{3}{3a}    &\adjoin&      \EBR{B}{3}{3a}             &  $\Gamma$--M \\   
    172.126 ($P6_41'$ -- II)&   \EBR{A}{3}{3a}    &\adjoin&      \EBR{B}{3}{3a}             &  $\Gamma$--M \textsuperscript{(1)}\\
    172.128 ($P_c6_4$ -- IV)&   \EBR{A}{6}{6a}    &\adjoin&      \EBR{B}{6}{6a}             &  $\Gamma$--M \textsuperscript{(1)}\\ \hline
    173.129 ($P6_3$ -- I)   &   \EBR{^1E}{2}{2a}    &\adjoin&    \EBR{^2E}{2}{2a}           &  $\Gamma$--K \\
    173.130 ($P6_31'$ -- II)&   \EBR{A_1}{2}{2a}    &\adjoin&    \EBR{^1E{}^2E}{4}{2a}      &  $\Gamma$--K \\    
    173.131 ($P6_3'$ -- III)&   \EBR{A_1}{2}{2a}    &\adjoin&    \EBR{^1E}{2}{2a}           &  \GKKprime \textsuperscript{(5)} \\   
    173.132 ($P_c6_3$ -- IV)&   \EBR{A_1}{2}{2a}    &\adjoin&    \EBR{^1E{}^2E}{4}{2a}      &  $\Gamma$--K \\
    174.135 ($P\bar{6}'$ -- III)&\EBR{A_1}{1}{1a}    &\adjoin&    \EBR{^1E{}^2E}{2}{1a}      &  \GKKprime \textsuperscript{(5)} \\   \hline  
    177.151 ($P6'2'2$ -- III)&   \EBR{A_1}{1}{1a}    &\adjoin&    \EBR{E}{2}{1a}  
    &  $\Gamma$--K, K--M \textsuperscript{(3)} \\ 
    182.181 ($P6_3'2'2$ -- III)& \EBR{A_1}{2}{2a}    &\adjoin&    \EBR{^1E}{2}{2a} 
    &  $\Gamma$--K, K--M \textsuperscript{(4)} \\
    183.185 ($P6mm$ -- I)  &    \EBR{A_1}{1}{1a}   &\adjoin&     \EBR{A_2}{1}{1a}        &  \allpoints \textsuperscript{(2)}\\ 
    183.186 ($P6mm1'$ -- II) &  \EBR{A_1}{1}{1a}   &\adjoin&     \EBR{A_2}{1}{1a}        &  \allpoints \textsuperscript{(2)}\\
    183.187 ($P6'm'm$ -- III) & \EBR{A_1}{1}{1a}   &\adjoin&     \EBR{A_2}{1}{1a}        &  \allprimed \textsuperscript{(5)} \\ \hline
    183.188 ($P6'mm'$ -- III) & \EBR{A_1}{1}{1a}   &\adjoin&     \EBR{A_2}{2}{2b}        &  \allpoints  \\
    183.189 ($P6m'm'$ -- III)  &\EBR{A}{1}{1a}   &\adjoin&     \EBR{^1E_2}{1}{1a}        &  \allpoints \\  
    183.190 ($P_c6mm$ -- IV)  & \EBR{A_1}{2}{2a}   &\adjoin&     \EBR{A_2}{2}{2a}        &  \allpoints \textsuperscript{(2)}\\      
    184.191 ($P6cc$ -- I)   &   \EBR{A}{2}{2a}   &\adjoin&     \EBR{^1E_2}{2}{2a}        &  \allpoints \textsuperscript{(2)}\\  
    184.192 ($P6cc1'$ -- II)  & \EBR{A}{2}{2a}   &\adjoin&     \EBR{^1E_2{}^2E_2}{4}{2a} &  \allpoints \textsuperscript{(2)}\\  \hline 
    184.193 ($P6'c'c$ -- III) & \EBR{A_1}{2}{2a}   &\adjoin&     \EBR{^1E{}^2E}{4}{2a} &  \GKKprime \textsuperscript{(5)} \\ 
    184.194 ($P6'cc'$ -- III) & \EBR{A_1}{2}{2a}   &\adjoin&     \EBR{^1E{}^2E}{4}{2a} &  $\Gamma$--K \\
    184.195 ($P6c'c'$ -- III) & \EBR{^1E_2}{2}{2a}   &\adjoin&     \EBR{^2E_1}{2}{2a} &  \allpoints \\  
    184.196 ($P_c6cc$ -- IV)  & \EBR{A}{2}{2a}   &\adjoin&     \EBR{^1E_2}{2}{2a} &  \allpoints \textsuperscript{(2)} \\
    185.197 ($P6_3cm$ -- I)   & \EBR{A_1}{2}{2a}   &\adjoin&     \EBR{A_2}{2}{2a}        &  \allpoints \textsuperscript{(2)}\\   \hline
    185.198 ($P6_3cm1'$ -- II)& \EBR{A_1}{2}{2a}   &\adjoin&     \EBR{A_2}{2}{2a}        &  \allpoints \textsuperscript{(2)}\\  
    185.199 ($P6_3'c'm$ -- III)&\EBR{A_1}{2}{2a}   &\adjoin&     \EBR{A_2}{2}{2a}        &  \allprimed \textsuperscript{(5)} \\  
    185.200 ($P6_3'cm'$ -- III)&\EBR{A_1}{2}{2a}   &\adjoin&     \EBR{^1E}{2}{2a}        &  $\Gamma$--K \\  
    185.201 ($P6_3c'm'$ -- III)&\EBR{A_1}{2}{2a}   &\adjoin&     \EBR{^1E}{2}{2a}        &  $\Gamma$--K \\ 
    185.202 ($P_c6_3cm$ -- IV)& \EBR{A_1}{2}{2a}   &\adjoin&     \EBR{A_2}{2}{2a}        &  \allpoints \textsuperscript{(2)}\\  \hline
    186.203 ($P6_3mc$ -- I) &   \EBR{A_1}{2}{2a}   &\adjoin&     \EBR{A_2}{2}{2b}        &  \allpoints \textsuperscript{(2)}\\   
    186.204 ($P6_3mc1'$ -- II)& \EBR{A_1}{2}{2a}   &\adjoin&     \EBR{A_2}{2}{2b}        &  \allpoints \textsuperscript{(2)}\\  
    186.205 ($P6_3'm'c$ -- III)&\EBR{A_1}{2}{2a}   &\adjoin&     \EBR{^1E}{2}{2a}        &  \GKKprime \textsuperscript{(5)} \\  
    186.206 ($P6_3'mc'$ -- III)&\EBR{A_1}{2}{2a}   &\adjoin&     \EBR{A_2}{2}{2b}        &  \allpoints \\   
    186.207 ($P6_3m'c'$ -- III)&\EBR{A_1}{2}{2a}   &\adjoin&     \EBR{^1E}{2}{2a}        &  $\Gamma$--K \\ \hline 
    186.208 ($P_c6_3mc$ -- IV)& \EBR{A_1}{2}{2a}   &\adjoin&     \EBR{A_2}{4}{4b}        &  \allpoints \textsuperscript{(2)}\\  
    187.211 ($P\bar{6}'m'2$ -- III)& \EBR{A_1}{1}{1a}   &\adjoin&     \EBR{E}{2}{1a}   
    &  $\Gamma$--K, K--M \textsuperscript{(3)} \textsuperscript{(4)} \\ 
    188.217 ($P\bar{6}'c'2$ -- III)& \EBR{A_1}{2}{2a}   &\adjoin&     \EBR{E}{4}{2a} 
    &  $\Gamma$--K, K--M \textsuperscript{(3)} \textsuperscript{(4)} \\
    189.223 ($P\bar{6}'2'm$ -- III)& \EBR{A_1}{1}{1a}   &\adjoin&     \EBR{A_2}{1}{1a}  &  \GKKprime, K--M, K'--M \textsuperscript{(3)} \textsuperscript{(5)} \\ 
    190.229 ($P\bar{6}'2'c$ -- III)& \EBR{A_1}{2}{2a}   &\adjoin&     \EBR{^1 E}{2}{2a}  &  \GKKprime, K--M, K'--M \textsuperscript{(3)} \textsuperscript{(4)} \textsuperscript{(5)} \\
    \hline \hline 
    \end{tabular}
    \caption{
    List of magnetic space groups compatible with RTP for spinless models. Remarks (1--5) indicated in the last column correspond to specifications detailed in the text of Sec.~\ref{sec:MSGs}.
    }
    \label{tab:RTP-spinless}
\end{table}

\begin{table}[t!]
    \centering
    \begin{tabular}{c|ccc|c} \hline \hline
	MSG number (label -- type)  &   \multicolumn{3}{c|}{example pair of EBRs}    & pairs of momenta supporting RTP \\ \hline 
    168.109 ($P6$ -- I)     &   \EBR{^1\overline{E}_3}{1}{1a}   &\adjoin& \EBR{^2\overline{E}_1}{1}{1a}  &  \allpoints \\
    168.110 ($P61'$ -- II)  &   \EBR{^1\overline{E}_1{}^2\overline{E}_1}{2}{1a}   &\adjoin& \EBR{^1\overline{E}_2{}^2\overline{E}_2}{2}{1a}   &  $\Gamma$--K \textsuperscript{(2)}\\
    168.111 ($P6'$ -- III)  &   \EBR{\overline{E}}{1}{1a}   &\adjoin& \EBR{^1\overline{E}{}^2\overline{E}}{2}{1a}         &  \GKKprime \textsuperscript{(5)} \\
    168.112 ($P_c6$ -- IV)  &   \EBR{^1\overline{E}_1}{2}{2a}   &\adjoin& \EBR{^2\overline{E}_2}{2}{2a}         &  $\Gamma$--K \textsuperscript{(2)}\\   
    171.121 ($P6_2$ -- I)   &   \EBR{^1\overline{E}}{3}{3a}    &\adjoin&      \EBR{^2\overline{E}}{3}{3a}       &  $\Gamma$--M \\ \hline
    172.125 ($P6_4$ -- I)   &   \EBR{^1\overline{E}}{3}{3a}    &\adjoin&      \EBR{^2\overline{E}}{3}{3a}       &  $\Gamma$--M \\
    173.129 ($P6_3$ -- I)   &   \EBR{^1\overline{E}}{2}{2a}    &\adjoin&    \EBR{^2\overline{E}}{2}{2a}         &  $\Gamma$--K \\
    173.130 ($P6_31'$ -- II)&   \EBR{\overline{EE}}{4}{2a}    &\adjoin&    \EBR{^1\overline{E}{}^2\overline{E}}{4}{2a}      &  $\Gamma$--K \\  
    173.131 ($P6_3'$ -- III)&   \EBR{\overline{E}}{2}{2a}    &\adjoin&    \EBR{^1\overline{E}}{2}{2a}      &  
    \GKKprime \textsuperscript{(5)} \\ 
    173.132 ($P_c6_3$ -- IV)&   \EBR{\overline{E}}{2}{2a}    &\adjoin&    \EBR{^1\overline{E}{}^2\overline{E}}{4}{2a}      &  $\Gamma$--K \\  \hline
    174.135 ($P\bar{6}'$ -- III)&   \EBR{\overline{E}}{1}{1a}    &\adjoin&    \EBR{^1\overline{E}{}^2\overline{E}}{2}{1a}      &  \GKKprime \textsuperscript{(5)} \\ 
    177.151 ($P6'2'2$ -- III)& 
    \EBR{^1\overline{E}}{1}{1a}    &\adjoin&    \EBR{\overline{E}_1}{2}{1a} 
    &  $\Gamma$--K, K--M \textsuperscript{(3)} \\
    182.181 ($P6_3'2'2$ -- III)&\EBR{\overline{E}}{2}{2a}    &\adjoin&    \EBR{^1\overline{E}}{2}{2a} 
    &  $\Gamma$--K, K--M \textsuperscript{(4)}\\
    183.185 ($P6mm$ -- I)  &    \EBR{\overline{E}_1}{2}{1a}   &\adjoin&     \EBR{\overline{E}_3}{2}{1a}        &  $\Gamma$--K \textsuperscript{(2)}\\ 
    183.186 ($P6mm1'$ -- II)  & \EBR{\overline{E}_1}{2}{1a}   &\adjoin&     \EBR{\overline{E}_3}{2}{1a}        &  $\Gamma$--K \textsuperscript{(2)}\\ \hline
    183.187 ($P6'm'm$ -- III)  & \EBR{^1\overline{E}}{1}{1a}   &\adjoin&     \EBR{^2\overline{E}}{1}{1a}        &  \allprimed \textsuperscript{(5)} \\
    183.188 ($P6'mm'$ -- III) & \EBR{^1\overline{E}}{1}{1a}   &\adjoin&     \EBR{^2\overline{E}}{2}{2b}        &  \allpoints \\
    183.189 ($P6m'm'$ -- III)  &\EBR{^1\overline{E}_2}{1}{1a}   &\adjoin&     \EBR{^2\overline{E}_1}{1}{1a}        &  \allpoints \\ 
    183.190 ($P_c6mm$ -- IV)  & \EBR{\overline{E}_1}{4}{2a}   &\adjoin&     \EBR{\overline{E}_3}{4}{2a}        &  $\Gamma$--K \textsuperscript{(2)}\\
    184.191 ($P6cc$ -- I)  &    \EBR{^1\overline{E}_1}{2}{2a}   &\adjoin&     \EBR{^1\overline{E}_2}{2}{2a}        &  $\Gamma$--K \textsuperscript{(2)}\\ \hline
    184.192 ($P6cc1'$ -- II)  & \EBR{^1\overline{E}_1{}^2\overline{E}_1}{4}{2a} &\adjoin& \EBR{^1\overline{E}_2{}^2\overline{E}_2}{4}{2a} & $\Gamma$--K \textsuperscript{(2)}\\ 
    184.193 ($P6'c'c$ -- III) & \EBR{\overline{E}}{2}{2a}   &\adjoin&     \EBR{^1\overline{E}{}^2\overline{E}}{4}{2a} &  \GKKprime \textsuperscript{(5)} \\ 
    184.194 ($P6'cc'$ -- III) & \EBR{\overline{E}}{2}{2a} &\adjoin& \EBR{^1\overline{E}{}^2\overline{E}}{4}{2a} & $\Gamma$--K \\
    184.195 ($P6c'c'$ -- III) & \EBR{^1\overline{E}_2}{2}{2a}   &\adjoin&     \EBR{^2\overline{E}_1}{2}{2a} &  \allpoints \\
    184.196 ($P_c6cc$ -- IV) & \EBR{^1\overline{E}_1}{2}{2a} &\adjoin& \EBR{^1\overline{E}_2}{2}{2a} & $\Gamma$--K \textsuperscript{(2)}\\ \hline
    185.197 ($P6_3cm$ -- I)  &  \EBR{^1\overline{E}}{2}{2a}   &\adjoin&     \EBR{\overline{E}_1}{4}{2a}        &  $\Gamma$--K \textsuperscript{(2)}\\ 
    185.198 ($P6_3cm1'$ -- II)& \EBR{^1\overline{E}{}^2\overline{E}}{4}{2a}   &\adjoin&     \EBR{\overline{E}_1}{4}{2a}        &  $\Gamma$--K \textsuperscript{(2)}\\
    185.199 ($P6_3'c'm$ -- III)&\EBR{^1\overline{E}}{2}{2a}   &\adjoin&     \EBR{^2\overline{E}}{2}{2a}        &  \allprimed \textsuperscript{(5)} \\
    185.200 ($P6_3'cm'$ -- III)&\EBR{\overline{E}}{2}{2a}   &\adjoin&   \EBR{^1\overline{E}}{2}{2a}     & $\Gamma$--K \\
    185.201 ($P6_3c'm'$ -- III)&\EBR{\overline{E}}{2}{2a}   &\adjoin&   \EBR{^1\overline{E}}{2}{2a}     & $\Gamma$--K \\ \hline
    185.202 ($P_c6_3cm$ -- IV) &\EBR{^1\overline{E}}{2}{2a}   &\adjoin&     \EBR{\overline{E}_1}{4}{2a}        &  $\Gamma$--K \textsuperscript{(2)}\\
    186.203 ($P6_3mc$ -- I) &   \EBR{^1\overline{E}}{2}{2a}   &\adjoin&     \EBR{\overline{E}_1}{4}{2a}        &  $\Gamma$--K \textsuperscript{(2)}\\
    186.204 ($P6_3mc1'$ -- II)& \EBR{^1\overline{E}{}^2\overline{E}}{4}{2a}   &\adjoin&   \EBR{\overline{E}_1}{4}{2a}        &  $\Gamma$--K \textsuperscript{(2)}\\
    186.205 ($P6_3'm'c$ -- III)&\EBR{^1\overline{E}}{2}{2a}   &\adjoin&   \EBR{\overline{E}}{2}{2a}        &  
    \GKKprime \textsuperscript{(5)} \\   
    186.206 ($P6_3'mc'$ -- III)&\EBR{^1\overline{E}}{2}{2a}   &\adjoin&   \EBR{^2\overline{E}}{2}{2b}        &  \allpoints \\ \hline
    186.207 ($P6_3m'c'$ -- III)&\EBR{^1\overline{E}}{2}{2a}   &\adjoin&   \EBR{\overline{E}}{2}{2a}        &  $\Gamma$--K \\ 
    186.208 ($P_c6_3mc$ -- IV)& \EBR{^1\overline{E}}{2}{2a}   &\adjoin&   \EBR{\overline{E}_1}{4}{2a}        &  $\Gamma$--K \textsuperscript{(2)}\\
    187.211 ($P\bar{6}'m'2$ -- III)& \EBR{^1\overline{E}}{1}{1a}   &\adjoin&     \EBR{\overline{E}_1}{2}{1a}   &  $\Gamma$--K, K--M \textsuperscript{(3)} \textsuperscript{(4)} \\ 
    188.217 ($P\bar{6}'c'2$ -- III)& \EBR{^1\overline{E}}{2}{2a}   &\adjoin&     \EBR{\overline{E}_1}{4}{2a}   &  $\Gamma$--K, K--M \textsuperscript{(3)} \textsuperscript{(4)} \\
    189.223 ($P\bar{6}'2'm$ -- III)& \EBR{^1\overline{E}}{1}{1a}   &\adjoin&     \EBR{^2\overline{E}}{1}{1a}  &  \GKKprime, K--M, K'--M \textsuperscript{(3)} \textsuperscript{(5)}  \\ \hline
    190.229 ($P\bar{6}'2'c$ -- III)& \EBR{\overline{E}}{2}{2a}   &\adjoin&     \EBR{^1 \overline{E}}{2}{2a}  &  \GKKprime, K--M, K'--M \textsuperscript{(3)} \textsuperscript{(4)} \textsuperscript{(5)} \\
    \hline \hline 
    \end{tabular}
    \caption{
    List of magnetic space groups compatible with RTP for spinful models. Remarks (2--5) indicated in the last column correspond to specifications detailed in the text of Sec.~\ref{sec:MSGs}.
    }
    \label{tab:RTP-spinful}
\end{table}

\noindent Additionally, we note MSG that contain $C_{2y}:(x,y,z,t)\mapsto (-x,y,-z,t)$ or $C_{2x}\mathcal{T}:(x,y,z,t)\mapsto(x,-y,-z,-t)$ symmetries and exhibit quantized polarization along the $\Gamma$--M line. This reduces the set of RTP-compatible lines to $\Gamma$--K and K--M, which then carry RTP of the same magnitude and opposite sign, $\Delta\mathscr{P}_{\Gamma,\textrm{K}} =- \Delta\mathscr{P}_{\textrm{K},\textrm{M}}$ [cf.~remark ``\textsuperscript{(4)}'' below].

For the hexagonal MSGs that are not discarded by the bullet points listed above, we identify those that can host RTP by the use of magnetic band representation tables of the Bilbao Crystallographic Server~\cite{elcoro2020magnetic, xu2020}. Namely, we find all MSGs that possess at least one pair of elementary band representations (EBRs) that exhibit disjoint 
symmetry indicators 
along at least a pair of high-symmetry lines in the rBZ.
The list of these MSGs for spinless (resp.~spinful) models with example elementary band representations (EBRs) and all RTP-supporting pairs of momenta in rBZ 
are presented in Tab.~\ref{tab:RTP-spinless} (resp.~Tab.~\ref{tab:RTP-spinful}).\\

\noindent Some entries in the tables are marked by one or more numbered labels in the last column; these indicate the following properties: 

\begin{itemize}
\item[\textsuperscript{(1)}] The lists of MSGs supporting RTP for spinless vs.~spinful models are not identical: there are four MSGs (e.g.~171.122) that support a pair of band representations obeying the mutually disjoint condition for the spinless but not for the spinful case. (MSGs supporting spinful but not spinless mutually disjoint EBRs do not exist.)

\item[\textsuperscript{(2)}] Some MSGs (e.g.~168.112) support RTP both for spinless and spinful models; however, the list of RTP-supporting pairs of momenta in rBZ for spinful models is reduced compared to the spinless case (The opposite disparity does not occur). 

\item[\textsuperscript{(3)}] In MSGs with $C_{2y}$ or $C_{2x}\mathcal{T}$ symmetries (e.g.~177.151) the polarization is constant along $\Gamma$--M line, i.e., this line does not support RTP. Hence, RTP along $\Gamma$--K and K--M are equal in magnitude and have opposite signs. 

\item[\textsuperscript{(4)}] For some MSGs with $C_{2y}$ or $C_{2x}\mathcal{T}$  (e.g.~182.181) we find EBRs with mutually-disjoint condition fulfilled only at $\Gamma$ and K, but not at $M$. According to comment (3), this is nonetheless sufficient to support an RTP along both $\Gamma$--K and K--M.

\item[\textsuperscript{(5)}] Some hexagonal MSGs do not possess the sixfold $C_{6z}$rotation (e.g.~ 168.111) and hence momenta K and $\tkpr$ are not related by symmetry. 
This allows to define independent RTPs for pairs of momenta that include either K or $\tkpr$.
\end{itemize}

The model in Eq.~(1) of the main text corresponds to MSG 183.189 ($P6m'm'$ -- type I).
We also remark that eight (in the spinless case) resp.~six (in the spinful case) of the listed magnetic space groups are of type II, i.e.~they are the ``usual'' non-magnetic space groups with time-reversal symmetry. These  instances provide a natural starting point to perform a high-throughput search for the delicate-topological RTP phase by sifting through the databases of known non-magnetic crystalline compounds.\\


\end{document}